\documentclass[english]{iopart}

\usepackage{graphicx}
\usepackage{url}
\usepackage{iopams} 

\expandafter\let\csname equation*\endcsname\relax
\expandafter\let\csname endequation*\endcsname\relax

\usepackage{amssymb}
\usepackage[latin1]{inputenc}
\usepackage{amsmath}
\usepackage{epsfig}
\newcounter{fig}
\begin{document}

\title{Symmetries of non-linear ODEs: lambda extensions of the Ising correlations}
      
\vskip .1cm 

\author{S. Boukraa$^\dag$, 
J.-M. Maillard$^\ddag$, }
\address{\dag LSA, IAESB,
  Universit\'e de Blida, Algeria}
\address{\ddag\ LPTMC, Sorbonne Universit\'e,  Tour 23, 5\`eme \'etage, case 121, \\
 4 Place Jussieu, 75252 Paris Cedex 05, France} 
\ead{maillard@lptmc.jussieu.fr, jean-marie.maillard@sorbonne-universite.fr, 
bkrsalah@yahoo.com}

\vskip .1cm

\begin{abstract}
  
This paper provides
several illustrations of the numerous remarkable properties of the lambda-extensions
 of the two-point correlation functions of the Ising model,
sheding some light on the non-linear ODEs of the Painlev\'e type. We first show that this concept
also exists  for the factors of the  two-point correlation functions focusing, for pedagogical reasons,
on two examples namely  $\, C(0,5)$  and $\, C(2,5)$ at $\, \nu \, = \, -k$. 
We then display, in a learn-by-example approach,
some of the puzzling properties and structures of these 
lambda-extensions: for an infinite set of  (algebraic) values of  $\, \lambda$ these power series become algebraic functions,
and for a finite set of (rational) values of lambda they become D-finite functions, more precisely polynomials (of different degrees)
in the complete elliptic integrals of the first and second kind $ \, K$ and $\, E$. For generic values
of  $\, \lambda$ these power series are not D-finite, they are differentially algebraic. For an infinite number
of other (rational) values of  $\, \lambda$  these power series are globally bounded series,
thus providing an example of an infinite number of  globally bounded differentially algebraic series. Finally, taking the example
of a product of two diagonal two-point correlation functions, we suggest that many more families of
non-linear ODEs of the Painlev\'e type remain to be discovered on the two-dimensional Ising model, as well as their
structures, and in particular their associated lambda extensions. 
The question of their possible reduction, after complicated transformations, to Okamoto sigma forms of Painlev\'e VI
remains an  extremely difficult challenge. 

\end{abstract}

\vskip .1cm


\noindent {\bf AMS Classification scheme numbers}: 34M55, 
47E05, 81Qxx, 32G34, 34Lxx, 34Mxx, 14Kxx 

\vskip .2cm


\vskip .2cm

\vskip .2cm

\today 

\vskip .2cm


\vskip .2cm

\section{Introduction:  linear versus non-linear symmetry representations}
\label{symmetry}
It is not necessary to underline the fundamental role played by the concept of symmetry in physics~\cite{Gross},
or in applied mathematics, and in the foundations for the fundamental theories of modern physics. 
Symmetries can correspond to continuous or discrete transformations, and are frequently amenable
to mathematical formulations such as group representations, with invariant
or covariant properties, non-trivial identities,  conservation laws, ...

Integrable models (in dynamical systems, lattice statistical mechanics, quantum field theory, solid state physics, enumerative combinatorics, ...)
play a selected role, since they correspond to situations where one has ``enough'' (possibly an infinite number of)
conserved quantities to solve the problem. We are not going to recall the techniques and tools introduced to achieve that goal
(Yang-Baxter equations, Bethe Ansatz, Lax pairs, Schlesinger systems~\cite{Artamonov}, ...) but we will rather focus on the linear and non-linear
differential equations emerging naturally in these problems\footnote[5]{The non-linear ODEs emerging in such an ``integrable'' framework
are highly selected: they have the (fixed critical) Painlev\'e property,
they have algebraic function solutions, etc ... This is in (strong) contrast with the generic non-linear ODEs
for which more numerical analysis  (investigation of the qualitative behavior of non linear ODEs, stability and boundness, ...) 
must be performed (see for instance~\cite{Tunc,Tunc2,Athanassov}).}, and on the corresponding symmetries of these ordinary differential equations.
To address that problem we will, for pedagogical reasons, focus on the analysis of the two-point correlation functions
of a fundamental integrable model, the two-dimensional Ising model~\cite{Importance}.

Some  two-point correlation functions  $\, C(M, N)$ of the two-dimensional Ising model
can be seen as solutions of {\em linear} differential equations
and,  {\em in the same time}, also as solutions of  {\em non-linear} differential equations,
namely Okamoto sigma-forms of Painlev\'e VI equations. The solutions of these last non-linear
ODEs naturally introduce one-parameter families of power series solutions, that are called
lambda-extensions of the two-point correlation functions. 

The  two-point correlation functions $ \, C(M,N)$
we will consider~\cite{bmm,bmcm} for the special case $\, \nu = \, -k$ (or in the isotropic case $\,  \nu = \, 1$), are {\em polynomial expressions}
of the complete elliptic integrals of the first and second kind $\, K$ and $\, E$: 
they are solutions of {\em linear} differential operators with polynomial coefficients, in other words they are {\em D-finite}, 
however, when introducing some well-suited log-derivative of these two-point correlation functions (see (\ref{defsigma}) below),
they are also solutions of  highly selected non-linear differential equations having the Painlev\'e property~\cite{cosgrove}, namely 
 Okamoto sigma-forms~\cite{okamoto} of Painlev\'e VI (see (\ref{eqnmodd}) below), in other words they are
{\em differentially algebraic}\footnote[2]{A differentially algebraic function~\cite{Tutte,Moore} is a function 
  $\, f(t)$ solution of a polynomial relation $\, P(t, \, f(t), \, f'(t), \, \cdots \, f^{(n)}(t)) \, = \, \, 0$,
where $\, f^{(n)}(t)$ denotes the $n$-th derivative of  $\, f(t)$ with respect to $\, t$.}. 
The two-point correlation functions $ \, C(M,N)$ have {\em in the same time}, a linear (D-finite) description
and a  {\em  non-linear}  (differentially algebraic) description ! The question of the analysis of the symmetries of these two
{\em linear and non-linear} ordinary differential equations, and of the  symmetries of their
solutions\footnote[9]{The symmetries of a differential equation and the symmetries of the solutions
 of the differential equation are two different concepts.} naturally pops out. 
It is crucial to note that the {\em  non-linear} ordinary differential equations for the  two-point correlation functions $ \, C(M,N)$
correspond to {\em one closed equation} (see  (\ref{eqnmodd}) below) where the two integers $\, M$ and $\, N$ are parameters in the equation.
In contrast the {\em  linear} differential equations  for the $ \, C(M,N)$ correspond to an infinite number of  linear differential equations 
of order (and degree and size) growing with the two integers $\, M$ and $\, N$. Each description (linear versus non-linear)
has its own advantages and disadvantages: an infinite number of differential operators to be discovered but they are simply linear, versus
one ($M$, $N$-dependent) equation encapsulating everything, but it is non-linear. The analysis of the symmetries of the linear differential operators
associated with  the  two-point correlation functions $ \, C(M,N)$ can, for instance, be performed considering the corresponding
differential Galois group. Actually we have seen in previous papers~\cite{SelectedGalois}  that the linear differential operators emerging in the
integrable models are systematically associated with selected differential Galois groups, the operators being 
homomorphic~\cite{SelectedGalois}  to their adjoint associated operators\footnote[1]{We even have this remarkable property with most of the  linear differential operators
  annihilating diagonals of rational functions~\cite{SelectedGalois}.}. In this Ising case, the  linear differential operators are
homoomorphic~\cite{Holonomy} to the
symmetric $\, N$-th power of the order-two  linear differential operator annihilating the complete elliptic integrals of the first or second kind $\, K$ and $\, E$.
Along this line, some mathematicians could argue that, if a differential Galois group approach of integrability
is probably natural, an extension of the concept of differential Galois group for {\em non-linear} ODEs is certainly
hopeless in general~\cite{Morales}.  They may even argue (see~\cite{Morales} in section 6.2) that, even if most of the people that work
in integrability consider the families of Painlev\'e transcendents~\cite{Handbook,PainlTracy} as integrable, their opinion is that, in general, they
are non integrable\footnote[3]{At least in the (narrow) Liouville sense~\cite{Stoyanova,Christov}.}.
Let us recall that  the sigma-form of Painlev\'e VI equations
(like (\ref{eqnmodd}) below), are {\em highly selected non-linear ODEs}:
they have the {\em fixed critical point property}~\cite{Ince,Bureau1,Bureau2}
({\em Painlev\'e property}) and  can be obtained from isomonodromic deformations of linear differential equations~\cite{Fuchs,Fuchs2},
which allows to see these non-linear ODEs as compatibility conditions of a linear Schlesinger system of PDEs. In that case one could imagine
to consider a differential Galois Theory for the underlying Schlesinger system. The purpose of this paper is not to build a
differential Galois Theory of Painlev\'e equations in order to discuss, from a very general mathematical viewpoint the ``symmetries'' of the
non-linear ODEs (like (\ref{eqnmodd}) below) emerging for the $\ C(M, \, N)$ Ising two-point correlation functions.
On the contrary, in a very pedagogical, learn-by-examples approach, we will display a large set of the  properties (symmetries ...)
of the  $\ C(M, \, N)$ two-point correlation functions, with a focus on the remarkable properties\footnote[9]{We must also mention
  the fact that the {\em lambda-extensions} of the two-point correlation functions $\, C(M, \, N)$ also verify quadratic 
 {\em  discrete} recursions~\cite{perk,mccoy3,Orrick}
  ({\em  lattice recursions} in the two integers $\, M$ and $\, N$), that can be seen as integrable lattice recursions.} of
the {\em lambda-extensions} solutions of
the sigma-form of Painlev\'e VI non-linear ODEs (like (\ref{eqnmodd}) below). For pedagogical reasons we will restrict to
$\, C(0, \, 5)$ and $\, C(2, \, 5)$. Then, taking an example
of  product of two diagonal two-point correlation functions, we will  suggest that many more families of
non-linear ODEs of the Painlev\'e type remain to be discovered on the two-dimensional Ising model, as well as their
structures, and in particular their associated lambda extensions. 
Finally, we will give additional comments and results providing an illustration of a set of remarkable,
and sometimes puzzling, properties of the lambda-extensions of the Ising two-point correlation functions.  

\vskip .2cm

\section{Recalls}
\label{introduction}

We revisit, with a pedagogical heuristic motivation, the lambda extensions~\cite{Holonomy} 
of some two-point  correlation functions $ \, C(M,N)$ of the two-dimensional
Ising model. For simplicity we will examine in detail the lambda extensions
of a particular low-temperature diagonal correlation function, namely $ \, C(0,5)$ and  $ \, C(2,5)$,
in order  to make crystal clear some structures and subtleties.
Note however that similar structures and results can also be obtained on other 
two-point  correlation functions  $ \, C(M,N)$ for the special case $\, \nu = \, -k$
studied in~\cite{bmm} where Okamoto sigma-forms of Painlev\'e VI
equations also emerge.

In a previous paper~\cite{bmm} we considered the two-point correlation $ \, C(M,N)$
of spins at sites $\, (0,0)$ and  $\, (M,N)$, 
of the anisotropic Ising model defined
by the interaction energy
\begin{eqnarray}
  \hspace{-0.7in} \quad  \quad  \quad \quad  \quad \quad 
         {\mathcal E \,}
\, \, = \, \,\, -\sum_{j,k}\{E_v\sigma_{j,k}\sigma_{j+1,k}
\, +E_h\sigma_{j,k}\sigma_{j,k+1}\}, 
\end{eqnarray}
where $ \, \sigma_{j,k}\, = \,\, \pm  \, 1 \, $ is the spin at row $\, j$
and column $\, k$,  and where the sum is over all lattice sites. 
Defining
\begin{eqnarray}
\hspace{-0.9in} 
  k\,\,= \,\, \,(\sinh 2E_v/k_BT \, \sinh 2E_h/k_BT)^{-1}
  \quad \quad \hbox{and} \quad \quad \quad
  \nu \, = \, \, \frac{\sinh 2E_h/k_BT}{\sinh 2E_v/k_BT}, 
\end{eqnarray}
we found~\cite{bmm} that in the special
case\footnote[5]{The condition
  $\, \nu \,  \,  = \, \, \, -k \, $ (as well as the isotropic case $\, \nu \, = \ 1$) is special
  because it is such that the complete elliptic integrals of the third kind (EllipticPi in Maple),
  appearing in the anisotropic case, 
  reduce to complete elliptic integrals of the second kind (see equation (30) in~\cite{bmm}).}
\begin{eqnarray}
  \quad  \quad  \quad \quad  \quad \quad  \quad
\nu \,  \,  = \, \, \, -k, 
\end{eqnarray}
the correlation\footnote[1]{Which is the same as the Toeplitz
  determinants~\cite{Comedy} of Forrester-Witte~\cite{fw} as given in~\cite{gil}.}
$ \, C(M,N)$ satisfies an Okamoto sigma-form of the 
Painlev{\'e} VI equation.

For $\, \, T\, <\, T_c$,  $\, M \, \le \,  N\, $ and $\, \,  \nu\, =\, -k$, 
with $\,\,  t=\, k^2$,  introducing
\begin{eqnarray}
  \label{defsigma}
\hspace{-.08in} \quad 
\sigma\,\, =\,\, \,
t \cdot \, (t-1) \cdot \, \frac{d \ln C(M,N)}{dt} \,\, \,-\frac{t}{4}, 
\end{eqnarray}
we have, when  $ \, M+N \, $ is odd,
the following Okamoto sigma-form of the 
Painlev{\'e} VI equation~\cite{bmm}:
\begin{eqnarray}
\label{eqnmodd}
&&\hspace{-.68in}
  t^2 \cdot \, (t-1)^2 \cdot \, \sigma''^2 \, \,  \,  \, 
  +4 \cdot \,  \sigma' \cdot \, (t\, \sigma'\,  -\sigma) \cdot \,
  \Bigl((t-1) \cdot \, \sigma'\,  -\sigma \Bigr)
\nonumber\\
&&\hspace{-.48in}
-M^2 \cdot \, (t\, \sigma' -\sigma)^2 \,\, \, \,  -N^2 \cdot \, \sigma'^2
\,\,   +(M^2 +N^2)
\cdot \,  \sigma' \cdot \, (t\, \sigma'\, -\sigma)
\, \,  \,  = \,  \, \, \, 0. 
\end{eqnarray}

\vskip .1cm

\subsection{Two factors}
\label{twofac}

In this $ \, M+N \, $ odd, $\, M \, \le \,  N$, $\, M \, \ne \, 0$, $\, \nu= \, -k \, $ case,
the correlation functions factor
into {\em two} factors\footnote[9]{What is written here in subsection \ref{twofac}
  is also true when $\, M= \, 0$, with the caveat that $\, g_{+}$ and $\, g_{-}$ in (\ref{twofactors}) 
factor into two factors (see subsection \ref{fourfac} below).}. 
We will write the factorizations of these $ \, C(M,N)$'s   as
\begin{eqnarray}
 \label{twofactors}
&& \hspace{-.46in} \quad 
 (1 \, -t)^{-1/4} \cdot \, C(M,N;t)  \,\,\, = \, \,\,\,\,  g_{+}(M,N;t)\cdot \, g_{-}(M,N;t), 
\end{eqnarray}
where the two factors $ \, g_{\pm}$ are {\em homogeneous polynomials} of the complete
elliptic integrals of the first and second kind:
\begin{eqnarray}
&& \hspace{-.36in} \quad \quad 
  {\tilde K}(k)  \, \, = \,  \,   \, \frac{2}{\pi} \cdot \, K(k) \,  \,= \, \,\, \, 
   {}_2F_1\Bigl([\frac{1}{2},\frac{1}{2}], \, [1], \,k^2\Bigr),
   \quad
  \nonumber\\
\label{elliptic}  
  &&\hspace{-.36in} \quad \quad 
  {\tilde E}(k) \, \, = \,  \,  \,  \frac{2}{\pi} \cdot \, E(k) \, \,  = \,  \,\, \, 
  {}_2F_1\Bigl(\frac{1}{2},-\frac{1}{2}], \, [1],  \,k^2\Bigr). 
\end{eqnarray}
We consider  the following logarithmic derivatives of
the previous two factors: 
\begin{eqnarray}
 \label{loggpm}  
\hspace{-.46in} \quad  \quad  \quad 
 \sigma_{\pm}(M,N;t) \, \,\,  = \, \, \,\,
  t \cdot \,(t-1) \cdot \, \frac{d\ln g_{\pm}(M,N;t)}{dt}.
\end{eqnarray}
The sigma functions have {\em additive} decompositions which follow from
the mulplicative decompositions  (\ref{twofactors}):
\begin{eqnarray}
\label{additive}
&&\hspace{-.56in} \quad \quad  \quad \quad 
 \, \sigma(M,N;t)\,\, = \, \,\,  \, \sigma_{+}(M,N;t)\,\, \, +\sigma_{-}(M,N;t). 
\end{eqnarray}
Here we begin with the factorizations (\ref{twofactors})  of  the $ \, \, C(M,N)$'s
with $ \, M+N \, $ odd, $\, M \, \le \, N$,
for miscellaneous values of $ \, M$ and $ \, N$, and,
by use of the methods described in~\cite{bmm} and of the
program {\em guessfunc} of Jay Pantone~\cite{pantone}, we find that
both $ \, \sigma_{+}(M,N;t)\,$ and $ \, \sigma_{-}(M,N;t)\,$ in (\ref{additive}) satisfy the
{\em same second-order non-linear differential equation}\footnote[5]{Note that this second order non-linear ODE,
  which is actually of the Painlev\'e type, is not of the Okamoto sigma-form of Painlev\'e VI form, but it can be reduced
to such a form using non trivial transformations (equations (26), (28) in section (2) of~\cite{bmcm}). }  
\begin{eqnarray}
  &&\hspace{-0.92in}\,\, 
 32 \,\, t^3 \cdot \, (t\, -1)^2 \cdot \, \sigma''^2
 \,\, \, + 4 \,\, t^2 \cdot \, (t-1) \cdot \,
\Bigl( 8 \cdot \, \sigma \, \, - 8 \cdot \,(t+1) \cdot \, \sigma'  \, \,
  +M^2 -N^2 \Bigr) \cdot \, \sigma'' 
\nonumber\\
&&\hspace{-0.9in} \, \,\, \,\, 
- \Bigl(8 \, \sigma \, \, -16 \cdot \, t \, \sigma' \,  +M^2 \, t \, -N^2 \, +1 \,  -t\Bigr)
\cdot
\Bigl(8 \cdot \, t \cdot \, (t-1) \cdot \, \sigma'^2 \,  \,
   -16 \, t \cdot \, \sigma  \cdot \, \sigma'
\nonumber\\
\label{nonlineareq}  
&&\hspace{-0.9in}
\quad   \quad   \quad \quad 
\, +8 \cdot \, \sigma^2 \, \, \, + (M^2-N^2) \cdot  \, \sigma \Bigr)
     \, \, \,  = \, \, \,\, \,  0, 
\end{eqnarray}
where the prime indicates a derivative with respect to $ \, t$, and where $\, \sigma$
is one of the two log-derivatives (\ref{loggpm}).

The two solutions (\ref{loggpm}) of (\ref{nonlineareq}),  $ \,  \, \sigma_{+}(M,N;t)\,$ and $ \, \sigma_{-}(M,N;t)$,
have different boundary conditions. 
Note that $\, \sigma_{\pm} \, = \, 0 \, \, $ {\em is a selected solution} of (\ref{nonlineareq}).  

\subsection{Four factors}
\label{fourfac}

In~\cite{bmm}, we discovered that $ \, C(0,N)$ with $ \, N$ odd and $ \, k= \, -\nu$, 
in the low-temperature regime, 
{\em factors into four terms} instead of two. The four factors for $ \, C(0,N)\, $
were presented as
\begin{eqnarray}
  \label{ffactors}
\hspace{-.3in}
  C(0,N) \, \, \, = \, \,  \,\, 
  {\rm constant}  \cdot  \, (1-t)^{1/2} \cdot  \, t^{(1-N^2)/4 }\cdot \,  f_1f_2f_3f_4, 
\end{eqnarray}
where the factors $ \, f_j \, $ all vanish at $ \,t= \, 0 \,$ in such a way to cancel
the factor $ \, t^{(1-N^2)/4}$.
We normalize the factors $\, f_i$ in (\ref{ffactors}) in such a way 
to extract a factor of $ \, (1-t)^{1/4} \, $
which is the limiting
behavior of $ \, C(0,N)$ as $ \, N \, \rightarrow \, \infty$,
and we impose the condition that {\em the four new factors satisfy the same non-linear differential equation}.
The previous factorization (\ref{ffactors}) in
{\em four} factors\footnote[1]{Examples of $\,  g_i(0,N)$'s
  for $ \, C(0,5)$ and $ \, C(0,7)$ are given in~\cite{bmcm}.}  now reads~\cite{bmcm}: 
\begin{eqnarray}
  \label{fourfactors}
\hspace{-.5in}
  (1\, -t)^{-1/4} \cdot \, C(0,N) \, \,   \, = \, \,  \, \,
  g_1(0,N) \cdot \,   g_2(0,N) \cdot  \,  g_3(0,N) \cdot \,  g_4(0,N).
\end{eqnarray}
If one defines
\begin{eqnarray}
  \label{sigmadef}
  \quad \quad \quad 
\sigma_j \, \,  = \, \, \, t  \cdot \, (t-1)  \cdot \, \frac{d\ln g_j(t)}{dt}, 
\end{eqnarray}
the previous factorization (\ref{fourfactors})  in four factors  becomes
an {\em additivity property} of the corresponding $\, \sigma_i$'s:
\begin{eqnarray}
\label{sigmaaddfour}
\hspace{-.8in} \quad \quad \quad
  \sigma(0, \, N) \, \, \, = \, \, \, \,\, 
   \sigma_1(0, \, N) \, \,  \,  +\sigma_2(0, \, N) \, \, \, 
      +\sigma_3(0, \, N) \, \, \,   +\sigma_4(0, \, N).  
\end{eqnarray}
These $\, \sigma_i$'s are solutions of the {\em same}  non-linear differential equation of the Painlev\'e type 
 which reads:
\begin{eqnarray}
  &&\hspace{-.99in} \,\,
t^2 \cdot \, (t \, -1)^2  \cdot \, \sigma''^2 \,\, \,  \,\,
+4\, \sigma' \cdot \, (t \cdot \, \sigma' \, -\sigma) \cdot \, \Bigl((t  -1) \cdot \, \sigma' \, -\sigma\Bigr)
\nonumber\\
&&\hspace{-.88in}
   +\frac{1}{4} \cdot \, \Bigl( (N^2+1) \cdot \, (t-1) \, -t^2 \Bigr)\cdot \, \sigma'^2 \,\,\,
   -\frac{1}{2^6} \cdot \,
   \Bigl( 16 \cdot \,  (N^2 \, +1 \, -2\, t ) \cdot \, \sigma  \, +N^2 \cdot \, t  \Bigr)\cdot \, \sigma'
   \nonumber\\
\label{4okamotoinsigma}
&&\hspace{-.55in} \quad  \quad 
   -\frac{1}{4} \cdot \, \sigma^2 \,  \, \,  \, +\frac{N^2}{2^6} \cdot \, \sigma \,  \,  \, \, 
 -{{ N^2 \cdot \, (N^2 \, -3) } \over { 2^{10}}} 
   \, \,  \, = \, \, \,  \,\, 0. 
\end{eqnarray}

\vskip .3cm 

\section{$\alpha$-extension of the four factors $ \, f_1$, $f_2$, $f_3$, $f_4 \, $ for $\, C(0, \, 5)$}
\label{muext}

We underlined that the (low-temperature) row correlation
functions  $\, C(0,N)$ factor, when is $\, N$ odd, into four factors (\ref{ffactors}).
These four factors $ \, f_i$'s are each a {\em homogeneous polynomial} of the complete
elliptic functions $\, E$ and $\, K$. Furthermore one can see
that  {\em each of these  four factors is a Toeplitz determinant}
(see for instance section G.4 of appendix G in~\cite{bmcm}). 

More specifically let us revisit  the $\, N=\, 5 \, $ case
detailed in~\cite{bmm} and also~\cite{bmcm},
where the two-point correlation $\, C(0,5) \, $ factors as follows
\begin{equation}  
 \label{C05}
\hspace{-.58in}
 C(0,5)\, \, = \, \,\, \, 
 {{256} \over {81 }}  \cdot \, {{ (1\, -t)^{1/2} } \over { t^6}}
 \cdot \, f_1 \cdot \, f_2 \cdot \, f_3 \cdot \, f_4, 
\end{equation}
where:
\begin{eqnarray}
\label{f1f2f3f4A}
&&\hspace{-.98in} \quad \,\, \,  
  f_1\,\, =\,\, \,
  (2t\, -1) \cdot \, {\tilde E}\,\, +(1\, -t) \cdot \, {\tilde K},
 \quad  \,\, \,\,\,\,
 f_2\,\, =\,\,\,
 (1\, +t) \cdot \, {\tilde E}\,\, -(1\, -t) \cdot \, {\tilde K},
\\
&&\hspace{-.98in} \quad \quad \quad  \quad \quad
\label{f1f2f3f4B}
f_3\, =\,\, \,  (t\, -2) \cdot \, {\tilde E}\,\, 
+2 \cdot \, (1\, -t) \cdot \, {\tilde K}, \, 
  \\
  &&\hspace{-.98in} \quad \quad \quad  \quad \quad
  \label{f1f2f3f4C}
f_4\, =\,\,\,
3 \, {\tilde E}^2\,\, +2 \cdot \, (t\, -2) \cdot \,{\tilde E}\, {\tilde K}
\,\,  +(1-t) \cdot \, {\tilde K}^2. 
\end{eqnarray}

These exact polynomial expressions in terms of complete elliptic integrals of the first and second
kind $\,  {\tilde K}$ and $\,  {\tilde E}$, {\em actually have some lambda-extensions}. Considering the non-linear ODE's
verified by these $\, f_n$'s one can,  by a down-to-earth, order by order expansion
of the analytic at $\, t \, = \,  0 \, $ solution, find the series expansion of a
{\em one parameter   family of solution} of the non-linear ODE's
(we will denote $\, \alpha$ this parameter),
such that $\, \alpha = \, 0 \, $ corresponds to the previous  exact expressions (\ref{f1f2f3f4A}),
(\ref{f1f2f3f4B}), (\ref{f1f2f3f4C}).
The first terms of these  $\, \alpha$-dependent solutions read:
\begin{eqnarray}
\label{f_1alpha}
&&\hspace{-.98in} \,\, \,  
f_1(\alpha) \,\, =\,\, \,\,
{{3} \over {2}} \, \,t \, \,\, -{\frac {9\,{t}^{2}}{16}} \, \,\,
-{\frac {15 \, {t}^{3}}{128}} \, \,\,
- \left( {\frac{105}{2048}}  +{\frac{15}{1024}} \, \alpha \right)   \cdot \,  {t}^{4}
\, \,
- \left( {\frac{945}{32768}} +{\frac { 135  }{8192}} \, \alpha  \right) \cdot \,  {t}^{5}
\nonumber \\
&&\hspace{-.98in} \quad  \quad \quad \,\, \,  
\, \, - \left( {\frac{4851}{262144}} +{\frac { 513 }{32768}} \, \alpha \right)  \cdot \, {t}^{6} \,\, \,  \,
- \left( {\frac{27027}{2097152}}   +{\frac {7497 }{ 524288}} \, \alpha\right)  \cdot \, {t}^{7}
 \\
&&\hspace{-.98in} \quad \quad  \quad \quad \,\, \,   \, \,
   - \left( {\frac{637065}{67108864}} \, +{\frac {434295 }{33554432}} \,\alpha\right) \cdot \,  {t}^{8}
   \nonumber \\
&&\hspace{-.98in} \quad \quad \quad \quad \quad \
\, \, - \left( {\frac{15643485}{2147483648}}   \, + {\frac {6292455 }{536870912 }} \  \,\alpha
\, \, -{\frac { 105 }{ 536870912}} \  \,\alpha^2 \right)  \cdot \, {t}^{9}
\, \, \,\, \, + \, \, \cdots \nonumber
\end{eqnarray}
\begin{eqnarray}
\label{f_2alpha}
&&\hspace{-.98in}  \,\, \,  
f_2(\alpha)\,\, =\,\, \,
{{3} \over {2}} \,t \, \,  -{\frac {3\,{t}^{2}}{16}} \,\,\, 
   -{\frac {3\,{t}^{3}}{128}}\,\,\, 
   - \left( {\frac{15}{2048}} - {\frac{15}{1024}} \, \alpha \right) \cdot \,  {t}^{4} \,\,\, 
  - \left( {\frac{105}{32768}} - {\frac {165 }{ 8192 }}\, \alpha \right)  \cdot \, {t}^{5}
  \nonumber \\
&&\hspace{-.98in} \quad \quad \quad \,\, \,  \,\,
  - \left( {\frac{441}{ 262144}} -{\frac {723 }{ 32768}} \,\alpha \right) \cdot \, {t}^{6} \,\,\, 
  - \left( {\frac{2079}{2097152}} -{\frac { 11799 }{524288}} \, \alpha\right) \cdot \,  {t}^{7}
  \\
&&\hspace{-.98in} \quad \quad \quad \,\, \,  \,\,
  - \left( {\frac{42471}{67108864}} \, -{\frac {747927  }{  33554432 }} \, \alpha \right)
  \cdot \, {t}^{8}
   \nonumber \\
&&\hspace{-.98in} \quad\quad \quad \quad \quad \quad  \,\,
   - \left( {\frac{920205}{2147483648}} \, -{\frac {11692785  }{ 536870912 }} \, \alpha \,
   -{\frac { 105 }{ 536870912 }} \, \alpha^2 
  \right) \cdot \, {t}^{9}
\, \,\,\, + \, \, \cdots \nonumber 
\end{eqnarray}
\begin{eqnarray}
\label{f_3alpha}
&&\hspace{-.98in} \quad \,\, \,  
f_3(\alpha)\,\, =\,\, \,
- {{3} \over {8}} \,{t}^{2} \,\,  \,  -{\frac {3\,{t}^{3}}{32}} \, \,\, 
-{\frac {45\,{t}^{4}}{1024}}\, \,\,   -{\frac {105\,{t}^{5}}{4096}} \, 
   \, \,  - \left( {\frac{2205}{131072}} -{\frac { 15}{ 131072}} \, \alpha \right) \cdot \, {t}^{6}
   \nonumber \\
&&\hspace{-.98in} \quad \quad \quad \quad \quad
\, \,  - \left( {\frac{6237}{524288}} -{\frac { 135 }{524288}} \, \alpha \right)\cdot \,  {t}^{7} \, \, 
   \, \,  - \left( {\frac{297297}{33554432}} -{\frac {3285 }{8388608 }}\, \alpha \right) \cdot \, {t}^{8}
   \nonumber \\
&&\hspace{-.98in} \quad \quad \quad \quad \quad \quad
  \, \, - \left( {\frac{920205}{134217728}} -{\frac { 16965 }{33554432}} \, \alpha\right) \cdot \, {t}^{9}
  \, \, \, \,  \, + \, \, \cdots 
\end{eqnarray}
\begin{eqnarray}
\label{f_4alpha}
&&\hspace{-.98in} \quad \,\, \,  
f_4(\alpha) \,\, =\,\, \,
- {{3} \over {8}}\,{t}^{2} \,\,  \,  -{{3} \over {16}}\,{t}^{3} \, 
 \, \,  -{\frac {129\,{t}^{4}}{1024}} \,  \, -{\frac {195\,{t}^{5}}{2048}} \,\,  \, 
 -\left( {\frac{5025}{65536}}  +{\frac { 15 }{131072 }} \, \alpha \right)
 \cdot \,  {t}^{6}
 \nonumber \\
&&\hspace{-.98in} \quad \quad \quad  \quad \quad \, \,
 - \left( {\frac{8421}{131072}} +{\frac {75 }{262144}} \, \alpha \right)
 \cdot \,  {t}^{7} \,\,  \,\,
 - \left( {\frac{1856253}{33554432}} +{\frac {3975  }{8388608}} \, \alpha \right)
 \cdot \, {t}^{8}
   \nonumber \\
&&\hspace{-.98in} \quad \quad \quad \quad \quad \quad \quad  \, \,\, \,
   - \left( {\frac{3260907}{67108864}} +{\frac {11025 }{16777216}} \right)
   \cdot \,  {t}^{9}
\, \,\, \,  \,  + \, \, \cdots 
\end{eqnarray}

Furthermore one sees, on the series expansions of the $\, \alpha$-extensions
(\ref{f_1alpha}),  (\ref{f_2alpha}),
(\ref{f_3alpha}),  (\ref{f_4alpha}), the following remarkable identities
\begin{eqnarray}
\label{identity}
&&\hspace{-.98in} \quad \quad \,\, \,
(1\, -t)^{1/4} \cdot f_2(\alpha)\, \, = \, \, f_1(1 \, -\alpha), \quad \quad  \, \,
(1\, -t)^{1/4} \cdot f_2(1 \, -\alpha)\, \, = \, \, f_1(\alpha),
 \nonumber \\
&&\hspace{-.98in} \quad  \quad   \,\, \,
 (1\, -t)^{1/4} \cdot f_4(\alpha)\, \, = \, \, f_3(1 \, -\alpha), \quad \quad  \, \,
 (1\, -t)^{1/4} \cdot f_4(1 \, -\alpha)\, \, = \, \, f_3(\alpha),
\end{eqnarray}
and thus:
\begin{eqnarray}
\label{identitythus}
&&\hspace{-.98in} \quad \quad \quad \quad  \quad \quad \,\, \,
(1\, -t)^{1/2} \cdot f_2(\alpha) \cdot f_4(\alpha)
\, \,\,  = \, \, \, \, f_1(1 \, -\alpha) \cdot f_3(1 \, -\alpha),
\nonumber \\
&&\hspace{-.98in} \quad  \quad  \quad \quad  \quad \quad  \,\, \,
(1\, -t)^{1/2} \cdot f_2(1 \, -\alpha) \cdot f_4(1 \, -\alpha)
\, \, \, = \, \, \, \, f_1(\alpha) \cdot f_3\alpha),
  \\
  &&\hspace{-.98in}   \,\, \, \quad  \quad \quad \quad \quad \quad
\label{identitythus2}
f_4(\alpha) \cdot f_1(1 \, -\alpha)
\, \, = \, \, \,  f_2(\alpha) \cdot f_3(1 \, -\alpha),
\quad  \, \,\nonumber \\
&&\hspace{-.98in}   \,\, \, \quad \quad \quad \quad  \quad \quad 
f_4(1\, -\alpha) \cdot f_1(\alpha)
\, \, = \, \, \,  f_2(1\, -\alpha) \cdot f_3(\alpha).  
\end{eqnarray}
In particular one has:
\begin{eqnarray}
\label{particular}
&&\hspace{-.98in}  \,\, \,  \quad \quad 
  f_1(0)\,\, =\,\, \,
   (2t\, -1) \cdot \, {\tilde E}\,\, +(1\, -t) \cdot \, {\tilde K},
\\
  &&\hspace{-.98in} \,\, \,\quad \quad 
   f_1(1)\,\, =\,\, \,
   (1\, -t)^{1/4} \cdot \,
   \Bigl((1\, +t) \cdot \, {\tilde E}\,\, -(1\, -t) \cdot \, {\tilde K}\Bigr), 
 \\
  &&\hspace{-.98in} \,\, \,\quad \quad 
 \label{particular2}
 f_2(0)\,\, =\,\,\,
     (1\, +t) \cdot \, {\tilde E}\,\, -(1\, -t) \cdot \, {\tilde K},
  \\
  &&\hspace{-.98in} \,\, \,\quad \quad 
     \label{particular21}
     f_2(1)\,\, =\,\,\, (1\, -t)^{-1/4} \cdot \,
     \Bigl( (2t\, -1) \cdot \, {\tilde E}\,\, +(1\, -t) \cdot \, {\tilde K} \Bigr), 
\\
&&\hspace{-.98in} \,\, \,  \quad \quad 
\label{particular3}
f_3(0)\, =\,\, \,  (t\, -2) \cdot \, {\tilde E}\,\, 
+2 \cdot \, (1\, -t) \cdot \, {\tilde K}, \, \\
&&\hspace{-.98in} \,\, \,  \quad \quad 
   \label{particular31}
   f_3(1)\, =\,\, \,
   (1\, -t)^{-1/4} \cdot \,
   \Bigl( 3 \, {\tilde E}^2\,\, +2 \cdot \, (t\, -2) \cdot \,{\tilde E}\, {\tilde K}
\,\,  +(1-t) \cdot \, {\tilde K}^2  \Bigr), 
  \\
  &&\hspace{-.98in}\,\, \,  \quad \quad 
  \label{particular4}
f_4(0)\, =\,\,\,
3 \, {\tilde E}^2\,\, +2 \cdot \, (t\, -2) \cdot \,{\tilde E}\, {\tilde K}
\,\,  +(1-t) \cdot \, {\tilde K}^2, \\
&&\hspace{-.98in} \,\, \,  \quad \quad 
 \label{particular41}
 f_4(1)\, =\,\, \,     (1\, -t)^{1/4} \cdot \,  \Bigl( (t\, -2) \cdot \, {\tilde E}\,\, 
+2 \cdot \, (1\, -t) \cdot \, {\tilde K}  \Bigr).
\end{eqnarray}
It is worth noticing that (in contrast with the $\, \lambda$-extension $\, C(0,5; \,\lambda)$  see (\ref{C05L}) below),
the $\, f_n(\alpha)$'s have {\em two} different values of the parameter $\, \alpha \, $ for which these
$\, \alpha$-extensions are D-finite, being (homogeneous) polynomials in $\, {\tilde E}$ and $\, {\tilde K}$. One
remarks with (\ref{particular3}) and (\ref{particular31}) (or (\ref{particular4}) and (\ref{particular41})),
that the corresponding  polynomials in  $\,  {\tilde E}$ and  $\,  {\tilde K}$ {\em are not necessarily of the same degree}
in $\,  {\tilde E}$ and $\,  {\tilde K}$.

The $\, \lambda$-extension  $\, C(0,5; \,\lambda )$
solution of the same non-linear ODE verified by $\, C(0,5)$ (namely (\ref{eqnmodd}) for $\, N= \, 5$) 
corresponds to the form-factor expansion~\cite{Holonomy,bm} which amounts to seeing this one-parameter family
of solutions as a deformation of the $\, (1\, -t)^{1/4} \, $ algebraic solution of the previous non-linear
ODE  (\ref{eqnmodd}) verified by $\, C(0,5)$: 
\begin{eqnarray}
  \label{C05L}
&&\hspace{-.98in}   \,\, \,  
   C(0,5; \,\lambda ) \,\, = \,\, \,\,  (1\, -t)^{1/4} \cdot
   \Bigl( 1 \, + \, \lambda^{2\, n} \cdot \, \sum_{n=1}^{\infty} \, f_{0,5}^{2\, n} \Bigr) 
 \\
  &&\hspace{-.98in} \quad \,\, = \,\, \,\, 
     1\, \,\,\, - {{t} \over {4}} \, \, \, \,-{\frac {3\,{t}^{2}}{32}}
\, \,\, -{\frac {7\,{t}^{3}}{128}}\, \,\,\,  -{\frac {77\,{t}^{4}}{2048}}\,\,  \,
-{\frac {231\,{t}^{5}}{8192}} \, \, -\left({{1463} \over {65536}} \,  +{{25} \over {1048576 }}
\cdot \,  \lambda^2\right) \cdot \, t^6
 \nonumber \\
  &&\hspace{-.98in} \quad  \quad
 \, -\left({{4807} \over {262144}} \, +{{275} \over {4194304}}
 \cdot \,  \lambda^2\right)
 \cdot \, t^7
 \, -\left({{129789} \over {8388608}} \, +{{123475} \over {1073741824}}
 \cdot \,  \lambda^2\right)
\cdot \, t^8
 \, \, \, + \, \, \cdots
\nonumber
\end{eqnarray}

\vskip .1cm

\subsection{Deformation around a  D-finite solution.}
\label{subDfinite}

The $\, \lambda$-extension of the correlation function $\, C(0,5; \,\lambda )$ can {\em also} be seen as a
$\, \mu$-deformation of the series of the correlation $\, C(0,5)$  which exact expression is given  by (\ref{C05})
(with (\ref{f1f2f3f4A}), (\ref{f1f2f3f4B}), (\ref{f1f2f3f4C}))
in terms of polynomials in  $\,  {\tilde E}$  and  $\,  {\tilde K}$. 
This one-parameter $\, \mu$-family of series expansion which verifies the same
non-linear ODE (\ref{eqnmodd}) as $\, C(0,5)$, reads:
\begin{eqnarray}
  \label{C05mu}
&&\hspace{-.98in}    \,  
  C(0,5; \,\lambda ) \,\, = \,\, \,\,  \,C(0,5) \, \,
  +  \mu \cdot \, G_1(t)  \,  \, + \mu^2 \cdot \, G_2(t)  \, \,  + \mu^3 \cdot \, G_3(t)
  \, \, \, + \, \, \, \cdots  \nonumber 
\, \, \\
&&\hspace{-.98in} \quad \,\, \,
 \, = \,\, \,\,\,  1\, \,\,\,  \,- {{t} \over {4}} \, \, \, \,-{\frac {3\,{t}^{2}}{32}} \,
\, \,\, -{\frac {7\,{t}^{3}}{128}}\, \,\,\,  \, -{\frac {77\,{t}^{4}}{2048}}\,\,  \,
-{\frac {231\,{t}^{5}}{8192}} \, - \left( {\frac{23433}{1048576}}    - {\frac{25}{1048576}} \,\mu \right)
\cdot \,  {t}^{6}  \nonumber 
\, \,  \\
&&\hspace{-.98in} \quad  \quad   \quad  \quad   \quad  
- \left( {\frac{77187}{4194304}} - {\frac{275}{4194304}} \,\mu \right)
\cdot \, {t}^{7}  \, \, - \left( {\frac{16736467}{1073741824}} \, -{  \frac { 123475 }{ 1073741824}} \, \mu \right)
   \cdot \, {t}^{8}
   \nonumber \\
&&\hspace{-.98in} \quad   \quad   \quad   \quad \quad  \quad   \quad     \quad  
  - \left( {\frac{57930653}{4294967296}} -{\frac { 708125}{4294967296 }} \, \mu \right)
 \cdot \,  {t}^{9}
\,\,\,\,\,  + \, \,\cdots 
\end{eqnarray}
The identification of these two power series (\ref{C05L}) and (\ref{C05mu})
corresponds to the simple relation between $\, \lambda$ and $\, \mu$:
\begin{eqnarray}
\label{lambdamu}
\hspace{-.58in} \quad \quad \quad \quad
  \lambda^2 \, \, = \, \,  \,  1 \, - \, \mu
  \quad \quad \quad \quad \hbox{or:}    \quad  \quad \quad \quad
  \mu  \, \, = \, \,  \,  1 \, - \,\lambda^2. 
\end{eqnarray}
This one-parameter series (\ref{C05L}), or (\ref{C05mu}), is
in agreement with a $\, \alpha$-extension
of the four products formula (\ref{C05})
\begin{equation}
  \label{C05lambda}
\hspace{-.58in} \quad \quad 
 C(0,5; \, \lambda) \, \, = \,\, \,\, \,  \, 
 {{256} \over {81 }}  \cdot \, {{ (1\, -t)^{1/2} } \over { t^6}}
 \cdot \, f_1(\alpha) \cdot \, f_2(\alpha) \cdot \, f_3(\alpha) \cdot \, f_4(\alpha), 
\end{equation}
if
\begin{eqnarray}
\label{relatalphaLambda}
&&\hspace{-.58in} \quad   \quad \,\, \,
   \mu \,\, \, = \, \, \,\, \,  4 \cdot \, \alpha \cdot \, (1 \, -\alpha)
   \quad   \quad \quad \hbox{or:}    \quad   \quad  \quad  \quad
   \lambda^2 \, \, = \, \,  \, \Bigl( 2 \, \alpha \, -1  \Bigr)^2, 
\end{eqnarray}
or:
\begin{eqnarray}
\label{relatalphaLambda-or}
&&\hspace{-.58in} \quad   \quad \quad   \quad   \quad   \quad   \quad \,\, \,
\alpha  \,\, \, = \, \, \,\, \,  {{ 1 \, \, \pm \, \lambda} \over { 2}}.
\end{eqnarray}
Thus one sees that the $\, \alpha \, \leftrightarrow \, 1 \, -\, \alpha \, \, $
involutive symmetry in the identities (\ref{identity})
amounts to changing the sign of $\, \lambda$:
$\,\, \lambda  \, \leftrightarrow \, -  \lambda$.
The value $\,\, \lambda  \, = \, 1$ (associated with the ``physical'' correlation functions)
corresponds to the two values $\, \alpha \, = \, 0$ and  $\, \alpha \, = \, 1$
for which the four factors $\, f_n$ become polynomials of $\,  {\tilde E}$  and  $\,  {\tilde K}$
(not necessarily of the same degree see for instance
(\ref{particular4}),  (\ref{particular41})).  The value $\,\, \lambda  \, = \, 0$
(associated with the algebraic function $\, C(0,5; \, 0) \, = \, \, (1\, -t)^{1/4}$)
corresponds to the  value $\,  \alpha \, = \, 1/2$.

Recalling the usual parametrization~\cite{bmcm,Holonomy} of the parameter
$\, \lambda$, namely $\, \lambda \, = \, \, \cos(u)$,
and the trigonometric identity
\begin{eqnarray}
\label{trigo}
  &&\hspace{-.58in} \quad \quad \quad \quad
     \cos(u) \,  \, = \, \,  \,  \,  2 \, \cos(u/2)^2  \,  \,  \, -1,   
\end{eqnarray}
we see that the parameter $\, \alpha \,$ is naturally parameterized as
\begin{eqnarray}
\label{alphaparam}
  &&\hspace{-.58in} \quad \quad \quad \quad  \quad   \quad
     \alpha \, \,\,   = \, \, \, \, \cos(u/2)^2, 
\end{eqnarray} 
the $\, \alpha \, \leftrightarrow \, 1 \, -\, \alpha \, $
involutive symmetry in the identities (\ref{identity}) 
corresponding to the parametrization
\begin{eqnarray}
\label{1minusalphaparam}
  &&\hspace{-.58in} \quad \quad \quad  \quad   \quad
1 \, - \, \alpha \, \, = \, \,\,  \, \,
1 \, \, - \, \cos(u/2)^2 \, \, \, = \, \, \, \, \sin( u/2)^2, 
\end{eqnarray} 
which amounts to changing $\, u$ into $\,\, u \, \rightarrow \, \, u \, + \, \pi \, $
in (\ref{alphaparam}),
a transformation that does not change  $\, \, \lambda^2 \, = \, \, \cos(u)^2$.

\vskip .2cm

\subsection{The algebraic $\alpha \, = \, \, 1/2$ case}
\label{alpha1/2}

One thus sees that the (involutive) symmetry
$\, \alpha \, \leftrightarrow \, \, 1 \, - \alpha \, $
 singles out  $\,  \alpha  \, = \, 1/2$. Along this line, note that, for $\, \alpha \, = \, 1/2$,
these $\, \alpha$-extensions
(\ref{f_1alpha}), (\ref{f_2alpha}) become {\em algebraic functions}. One actually has: 
\begin{eqnarray}
\label{f_1alpha1/2}
&&\hspace{-.98in} \,\,\quad \quad \,  
f_1\Bigl({{1} \over {2}} \Bigr) \,\, = \,\, \,\,
{{3} \over {2}} \, \cdot \, t \cdot \, (1-t)^{1/16}  \cdot \,
\Bigl( {{ 1 \, +(1-t)^{1/2}} \over {2 }}   \Bigr)^{5/4} 
 \\
&&\hspace{-.98in} \, \,   \quad \quad \quad \quad
\, = \,\, \,
{{3} \over {2}} \,  \,t \, \, -{\frac {9}{16}} \, t^2 \, \,\,
-{\frac {15}{128}} \, t^3  \,\,- {{15} \over {256 }}\,t^4 \, -{{1215} \over {32768 }}\,t^5
\, \, -{{6903} \over {262144}}\, t^6 \, \,\,
 \,  \,  + \, \, \cdots \nonumber 
\end{eqnarray}
\begin{eqnarray}
\label{f_2alpha1/2}
&&\hspace{-.98in} \,\,\quad \quad \,  
f_2\Bigl({{1} \over {2}} \Bigr) \,\, = \,\, \,\,
{{3} \over {2}} \, \cdot \, t \cdot \, (1-t)^{1/16}  \cdot \, (1-t)^{-1/4}  \cdot \,
\Bigl( {{ 1 \, +(1-t)^{1/2}} \over {2 }}   \Bigr)^{5/4} 
 \\
&&\hspace{-.98in} \quad  \quad \quad \quad  \quad
\, = \,\, \, \,  {{3} \over {2}} \,  \,t \, \,  \, -{\frac {3}{16}} \, t^2 \, \,\,
-{\frac {3}{128}} \, t^3  \,\,+ {{225} \over {32768 }}\,t^5 \, 
   \, \, +{{2451} \over {262144 }}\, t^6 \, \, \,  \,+ \, \, \cdots
   \nonumber 
\end{eqnarray}
The $\, \alpha$-extensions
(\ref{f_3alpha}), (\ref{f_4alpha}) for $\, f_3(\alpha)$ and  $\, f_4(\alpha)$
also become algebraic functions:
\begin{eqnarray}
\label{f_3alpha1/2}
&&\hspace{-.98in} 
f_3\Bigl({{1} \over {2}} \Bigr) \,\, = \,\, \,\, \, -{{3} \over {8}} \,  \,t^2
\, \, -{\frac {3}{32}} \, t^3 \, \,
-{\frac {45}{1024}} \, t^4  \,\,- {{105} \over {4096 }}\,t^5 \,\, 
   - {{4395} \over {262144 }}\,t^6 \,\, \,\, + \, \, \cdots
   \nonumber 
 \\
&&\hspace{-.98in} \quad   \, = \,\, 
-{{3} \over {8}} \cdot \, t^2 \cdot \, (1\, -t)^{1/16} \cdot \,
\Bigl( {{ 1 \, +(1-t)^{1/2}} \over {2 }}   \Bigr)^{-3/4} 
   \cdot \, \Bigl( {{(1\, +t^{1/2})^{1/2} \, \, -  (1\, -t^{1/2})^{1/2}} \over { t^{1/2}}} \Bigr)
   \nonumber
  \\
  &&\hspace{-.98in} \quad   \, = \,\,
-{{3} \over {8}} \cdot \, t^2 \cdot \, (1\, -t)^{1/16} \cdot \,
\Bigl( {{ 1 \, +(1-t)^{1/2}} \over {2 }}   \Bigr)^{-3/4} \cdot \,
\Bigl( 2 \cdot \, {{ ( 1\, -(1\, -t)^{1/2}) } \over {t}}  \Bigr)^{1/2}, 
\end{eqnarray}
\begin{eqnarray}
\label{f_4alpha1/2}
&&\hspace{-.98in} 
f_4\Bigl({{1} \over {2}} \Bigr) \,\, = \,\, \,\, \, -{{3} \over {8}} \,  \,t^2
\, \, -{\frac {3}{16}} \, t^3 \, \,
-{\frac {129}{1024}} \, t^4  \,\,- {{195} \over {2048 }}\,t^5 \,\, 
   - {{ 20115} \over {262144 }}\, t^6  \, \,  \,  \, + \, \, \cdots
   \nonumber 
 \\
&&\hspace{-.98in}  \quad \quad  \quad    \, = \,\, \,\, 
-{{3} \over {8}} \cdot \, t^2 \cdot \, (1\, -t)^{1/16} \cdot \,(1\, -t)^{-1/4} \cdot \,
\Bigl( {{ 1 \, +(1-t)^{1/2}} \over {2 }}   \Bigr)^{-3/4}
\nonumber \\
&&\hspace{-.98in} \quad \quad \quad   \quad  \quad \quad \quad \quad \quad
   \times \, \Bigl( {{(1\, +t^{1/2})^{1/2} \, \, -  (1\, -t^{1/2})^{1/2}} \over { t^{1/2}}} \Bigr)
   \nonumber 
\\
&&\hspace{-.98in}  \quad \quad  \quad   \, = \,\, \,\, 
-{{3} \over {8}} \cdot \, t^2 \cdot \, (1\, -t)^{1/16} \cdot \,(1\, -t)^{-1/4} \cdot \,
   \Bigl( {{ 1 \, +(1-t)^{1/2}} \over {2 }}   \Bigr)^{-3/4}
   \nonumber \\
&&\hspace{-.98in} \quad \quad \quad   \quad \quad  \quad \quad \quad  \quad  \quad \quad \quad
   \times \,   \Bigl( 2 \cdot \, {{ ( 1\, -(1\, -t)^{1/2}) } \over {t}}  \Bigr)^{1/2}.
\end{eqnarray}
One verifies easily that
\begin{eqnarray}
\label{verif}
  &&\hspace{-.98in} \quad \,\, \,\, \,
f_1\Bigl({{1} \over {2}} \Bigr) \cdot  f_3\Bigl({{1} \over {2}} \Bigr)
\,\, = \,\, \,\, \,  (1\, -t)^{1/2} \cdot \, f_2\Bigl({{1} \over {2}} \Bigr) \cdot f_4\Bigl({{1} \over {2}} \Bigr)
 \, \, = \, \, \,  - {{9} \over {16}} \cdot \, t^3 \cdot \,  (1\, -t)^{1/8},
  \\
  &&\hspace{-.98in}  \quad \,\, \,\, \,
     \label{verif2}
     f_1\Bigl({{1} \over {2}} \Bigr) \cdot  f_4\Bigl({{1} \over {2}} \Bigr)
     \,\, = \,\, \,\, \,  f_2\Bigl({{1} \over {2}} \Bigr) \cdot f_3\Bigl({{1} \over {2}} \Bigr)
   \, \, = \, \, \, \,
   - {{9} \over {16}} \cdot \, t^3 \cdot \,  (1\, -t)^{-1/8},
\end{eqnarray}
in agreement with the identities (\ref{identitythus}) and (\ref{identitythus2}).

\vskip .2cm

Do note that $\, f_1(\alpha)\,$ and $\, (1\, -t)^{1/4} \cdot f_2(\alpha)$,  but also
$\, t^{1/4} \cdot \, f_3(\alpha)\,$ and also $\, t^{1/4} \cdot \, (1\, -t)^{1/4} \cdot f_4(\alpha)$,
{\em verify the same Okamoto sigma-form of Painlev\'e VI} (independently of $\, \alpha$). 
Note that the previous algebraic function solutions  (\ref{f_1alpha1/2}) and   (\ref{f_2alpha1/2}) 
are actually such that  $\, f_1({{1} \over {2}})\,$ and $\, (1\, -t)^{1/4} \cdot f_2({{1} \over {2}})\,$
are not only solutions of the same non-linear ODE
but are actually the {\em same} algebraic function
$\, f_1({{1} \over {2}})\, =\, (1\, -t)^{1/4} \cdot f_2({{1} \over {2}})\,$.
Similarly  (\ref{f_3alpha1/2}) and   (\ref{f_4alpha1/2}) 
are actually such that  $\, f_3({{1} \over {2}})\,$ and $\, (1\, -t)^{1/4} \cdot f_4({{1} \over {2}})\,$
are not only solutions of the same non-linear ODE
but are actually the {\em same} algebraic function
$\, f_3({{1} \over {2}})\, =\, (1\, -t)^{1/4} \cdot f_4({{1} \over {2}})\,$.
For $\, \alpha \, = \, 1/2$ the corresponding $\, \lambda$ deduced from (\ref{relatalphaLambda})
is $\, \lambda \, = \, \, 0$ and the four product formula (\ref{C05lambda}) becomes, with the previous
exact algebraic expressions (\ref{f_1alpha1/2}), (\ref{f_2alpha1/2}),
(\ref{f_3alpha1/2}) and  (\ref{f_4alpha1/2}) (and after simplifications):
\begin{eqnarray}
 \label{C05lambdaalpha1/2}
&&\hspace{-.98in} \quad
 C(0,5; \, 0) \, \, = \, \,\, \,  \, 
 {{256} \over {81 }}  \cdot \, {{ (1\, -t)^{1/2} } \over { t^6}}
 \cdot \, f_1\Bigl({{1} \over {2}}\Bigr) \cdot \, f_2\Bigl({{1} \over {2}}\Bigr)\cdot  \,
 f_3\Bigl({{1} \over {2}}\Bigr) \cdot  \, f_4\Bigl({{1} \over {2}}\Bigr)
 \, \, = \, \,\, \,  \, (1\, -t)^{1/4}
 \nonumber \\
&&\hspace{-.98in} \quad  \,  \, \, \,  \,
\, \, = \, \,\, \,  \, 1 \,\, -{{1 } \over {4}} \, t \,\,
-{{3 } \over {32}} \, t^2 \,\, -{{7 } \over {128}} \, t^3\,
 \, -{{77 } \over {2048}} \, t^4 \,\, -{{231 } \over {8192}} \, t^5\,
 -{{1463 } \over {65536}} \, t^6\,
 \,  \, \, + \, \, \, \cdots 
\end{eqnarray}
in agreement with the $\, \lambda = \, 0$ evaluation of the form factor expansion (\ref{C05L}).
Note that, conversely, the identity (\ref{C05lambdaalpha1/2}) can be used to find the 
exact expressions of the products $\, f_1 \, f_4 \, $  and $\, f_1 \, f_3 \, $
evaluated at $\, \alpha \, = \, 1/2$
(see (\ref{verif}) and  (\ref{verif2})), when the exact expressions
on the $\, f_n$'s, $\, n \, = \, 1,\, 2,\, 3, \,4$,
are much more involved (see (\ref{f_1alpha1/2}), (\ref{f_2alpha1/2}),
(\ref{f_3alpha1/2}), (\ref{f_4alpha1/2})).

\vskip .2cm

{\bf Remark:} 
All these calculations are {\em not} specific of $\, N= \, 5$. Similar calculations
can be performed for other values of $\, N$. Since these calculations become more and more
involved, they will not be detailed here. Let us just give the expressions\footnote[1]{These expressions can be compared with
expressions (E.2) and (E.13) in appendix E of~\cite{bmcm} but with a different normalization (E.1).} of $\, f_1$ for different
(odd) values of $\, N$, in terms of the complete elliptic integrals of the first and second kind
 $\,  {\tilde K}$  and  $\,  {\tilde E}$ . 

\vskip .2cm

For $\, N=5, \, 7, \, 9,  \,\,$ the $\, f_1(N)$ solutions  read respectively:    
\begin{eqnarray}
\label{f1_5}
&& \hspace{-.9in}  
f_1(N=5) \, \, = \, \, \, (2\, t \, -1) \cdot \, {\tilde E} \, +(1\,-t) \cdot \, {\tilde K}, 
\end{eqnarray}
\begin{eqnarray}
\label{f1_7}
&& \hspace{-.9in} 
f_1(N=7) \, \, = \, \, \,
-(3\, t+4)  \cdot \,(t-1)^2  \cdot \, {\tilde K}^2 \,\,\,
+2\, (t \, -1) \cdot \, (3\, t^2-7\, t-4) \cdot \, {\tilde E}\, {\tilde K}
\nonumber \\
&& \hspace{-.9in}  \quad \quad \quad
\quad \quad \quad \quad \, +(11\, t^2-11\, t-4) \cdot \, {\tilde E}^2, 
\end{eqnarray}
\begin{eqnarray}
\label{f1_9}
&& \hspace{-.9in}  
f_1(N=9) \, \, = \, \, \,  (8\, t^2-47\, t+12) \cdot \, (t-1)^2 \cdot \, {\tilde K}^2
\nonumber \\
&& \hspace{-.9in}  \quad \quad \quad
\quad \quad \quad \quad\,
-2 \cdot \, (t-1) \cdot \, (16\, t^3-63\, t^2+83\, t-12) \cdot \, {\tilde E} \, {\tilde K}
\nonumber \\
&& \hspace{-.9in}  \quad \quad \quad
\quad \quad \quad \quad\,
+(32\, t^4-64\, t^3+151\, t^2-119\, t+12) \cdot \, {\tilde E}^2, 
\end{eqnarray}
We can verify for $\, N\, = \,5, \, 9, \, 13, \, \cdots \, $ that the factor  $\, f_1(N)\, $  expands as
\begin{eqnarray}
\label{f1N}
&& \hspace{-.9in}  \quad \quad \quad \quad \quad \quad  \quad   \quad 
f_1(N) \, \, = \, \, \,
\lambda_N \cdot \, t^{(N-1)^2/16} \, \,\, \, + \,\,\cdots,  
\end{eqnarray}
when, for $\, N\, = \,7, \, 11, \, 15, \, \cdots \, $ the factor  $\, f_1(N)$ has the expansion:
\begin{eqnarray}
\label{f1N}
&& \hspace{-.9in}  \quad \quad \quad \quad \quad  \quad \quad  \quad  \quad 
 f_1(N) \, \, = \, \, \,  \mu_N \cdot \, t^{(N+1)^2/16} \,\,\, + \,\,\cdots 
\end{eqnarray}

\vskip .2cm 

\subsection{Form-factor deformation around the algebraic function $\, f_1(1/2)$}
\label{formfactorf1}

Introducing a form-factor $\, \beta$-deformation around the algebraic function (\ref{f_1alpha1/2})
($\beta$ is the deformation parameter around $\, \alpha=\, 1/2$)
\begin{eqnarray}
\label{deduces2betaf1a}
  && \hspace{-.9in}   \quad  \quad 
f_1\Bigl({{1} \over {2}}\, + \, \beta\Bigr)
\, \, \, = \, \, \,
\nonumber \\
 && \hspace{-.9in}   \quad \quad \quad \quad \quad 
   {{3} \over {2}} \cdot \, t \cdot \, (1\, -t)^{1/16}  \cdot \,
   \Bigl( {{ 1 \, +(1\, -t)^{1/2}} \over {2}}  \Bigr)^{5/4}
   \, \, \, \, \, + \beta \cdot \, G(t)
   \, \,  \,  \, + \, \, \, \cdots 
\end{eqnarray}
and inserting (\ref{deduces2betaf1a}) in the non-linear ODE verified by
(\ref{deduces2betaf1a}), one gets an {\em order-three linear differential operator}
for the first coefficient $\, G(t)$.

This order-three linear differential operator has the following solution:
\begin{eqnarray}
\label{deduces2beta1f1asol}
&& \hspace{-.98in} 
G(t) \, \, = \, \, \,  \,  {{t^2} \over {64}}  \cdot \, (1\, -t)^{1/16}  \cdot \,
   \Bigl( {{ 1 \, +(1\, -t)^{1/2}} \over {2}}  \Bigr)^{1/4} \cdot \,
P_{E,K}  
\\
&& \hspace{-.98in}  \quad 
\, \, = \, \, \,\, 
-{{15} \over { 1024}} \cdot \, t^4 \, \,  -{{135} \over {8192}}  \cdot \, t^5 \, \, 
-{{513} \over {32768}}  \cdot \, t^6 \, 
\, -{{7497} \over {524288}}  \cdot \, t^7 \, \, 
-{{434295} \over {33554432}}  \cdot \, t^{8} \, 
\,\, + \,\,\, \cdots
\nonumber 
\end{eqnarray}
where $\, P_{E,K} \, \, $ is a {\em polynomial in $\, {\tilde E}$ and $\, {\tilde K}$}: 
\begin{eqnarray}
\label{deduces2betawheref1a}
&& \hspace{-.9in} \quad 
P_{E,K} \, \, = \, \, \,
 \\
&& \hspace{-.9in} \quad \quad \quad 
\Bigl(t-4 \, \,  +12 \cdot \, (1-t)^{1/2} \Bigr) \cdot \,
_2F_1\Bigl([{{3} \over {2}}, {{3} \over {2}}],[3],t\Bigr) \, \,\,\,
-8 \cdot \, _2F_1\Bigl([{{1} \over {2}}, {{1} \over {2}}],[2],t\Bigr)
\,\, \,\, = 
\nonumber \\
&& \hspace{-.9in} 
-8 \cdot \,
{{ 12 \cdot \, (t\, -2) \cdot \, (1 \, -t)^{1/2} \, + 3 \, t^2 \, -8\, t \, +8  } \over {t^2 }}
\cdot \,  {\tilde K}
\,\,\, -32 \cdot \, {{  t-2 +6 \cdot \, (1 \, -t)^{1/2}  } \over { t^2}}
\cdot \, {\tilde E}.
\nonumber
\end{eqnarray}
As far as the log-derivative with respect to $\, t$  is concerned, one gets:
\begin{eqnarray}
\label{deduces2betalog}
 && \hspace{-.98in}  
t \cdot \, (t \, -1) \cdot \, {{ d } \over {dt}} \ln\Bigl(f_1 \Bigl({{1} \over {2}} \, + \beta  \Bigr) \Bigr) 
\, \, \, = \, \, \, \,
 \, {{ 10 \cdot \, (1\, -t)^{1/2} \, +27 \, t \, -26  } \over {16 }} 
 \\
  && \hspace{-.98in}  \quad 
 - \, \, {{\beta} \over {96}}   \cdot \, 
 \Bigl( t \cdot \,  (1\, -t)^{1/2}\cdot \,   P_{E,K} \, \,
 + 2 \cdot \, t \cdot \, (1 \, -t) \, \cdot \, (1\, - \, (1\, -t)^{1/2})
 \cdot \,   {{ d P_{E,K}} \over {dt}}  \Bigr)
   \, \, + \, \, \, \cdots \nonumber
\end{eqnarray}
where the first deformation term is
also a polynomial in $\, {\tilde E}$ and $\, {\tilde K}$.

\vskip .2cm 

\section{$\alpha$-extensions of the two factors $F_1$, $F_2$ for $\, C(2, \, 5)$}
\label{muext}

The low-temperature correlation functions $\, C(M, \, N)$, at $\, \nu \, = \, -k$,
with $\, M \, < N$, $\, M+N$ odd, $\, M$ even but different from $\, 0$,
factor into the product of,
not four terms, but only two terms:
\begin{eqnarray}
\label{factor2int}
&& \hspace{-.9in} \quad \quad \quad \, \,\,
C(M, \,  N) \,\, = \, \, \,\,
\rho \cdot \, (1 \, -t)^{1/2} \cdot \,  t^{ -(N^2 -1)/4} \cdot \,
F_1(M,N) \cdot \, F_2(M,N).
\end{eqnarray}

For instance for $\, M \, = \, 2 \, $ and $\, N\, = \, \, 5 \,$ one has
\begin{eqnarray}
\label{factor2int25first}
&& \hspace{-.9in} \quad \quad \quad \quad \quad \, \,\,
C(2, \,  5) \,\, \, = \, \,\,\, 
{{ 256} \over { 2025 }} \cdot \, {{(1 \, -t)^{1/2} } \over { t^6 }} \cdot \,
F_1(2,5) \cdot \, F_2(2,5),
\end{eqnarray}
where
\begin{eqnarray}
\label{factor2int25F1}
&& \hspace{-.9in}
 \quad \quad  \, \,\,
F_1(2,5) \, \, = \, \,\, \,
2 \cdot \,(1 \, -t) \cdot \, (2\,t \, +1) \cdot \, {\tilde K}^2 \, \,
+(7\,t^2 \, -15\,t \, -4) \cdot \, {\tilde E} \, {\tilde K}  \, \,
\nonumber \\ 
&& \hspace{-.9in}
 \quad \quad \quad \quad \quad \quad \quad \quad \, \,\,
+(2\,t^2 \, +13\, t \, +2) \cdot \, {\tilde E}^2, 
\end{eqnarray}
and:
\begin{eqnarray}
\label{factor2int25F2}
&& \hspace{-.9in}
 \quad \quad  \, \,\,
F_2(2,5) \, \, = \, \,\, \,5 \cdot \, (t-1)^3 \cdot \,  {\tilde K}^3 \,
\, \, -(11\,t \, -17) \cdot \, (t \, -1)^2 \cdot \,  {\tilde E}\, {\tilde K}^2
\nonumber \\ 
&& \hspace{-.9in}
 \quad \quad  \quad  \quad \quad\, \,\,\,
+(t \, -1) \cdot \,(2\,t^2 \, -33\, t \, +19) \cdot \, {\tilde E}^2 \,  {\tilde K}
\, \,\, \,\, +(7\, t^2 \, -22\, t \, +7) \cdot \,  {\tilde E}^3. 
\end{eqnarray}
The $\, \lambda$-extension $\, C(2,5; \,\lambda )$ corresponds to a form-factor
expansion around the algebraic solution $\, (1\, -t)^{1/4}$: 
\begin{eqnarray}
  \label{C25L}
&&\hspace{-.98in}   \, 
   C(2,5; \,\lambda ) \,\, = \,\, \,\,  (1\, -t)^{1/4} \cdot
   \Bigl( 1 \, + \, \lambda^{2\, n} \cdot \, \sum_{n=1}^{\infty} \, f_{0,5}^{2\, n} \Bigr) 
 \\
  &&\hspace{-.98in} \quad  = \,\, \,\, 
     1\, \,\,\, - {{t} \over {4}} \, \, \, \,-{\frac {3\,{t}^{2}}{32}}
\, \,\, -{\frac {7\,{t}^{3}}{128}}\, \,\,\,  -{\frac {77\,{t}^{4}}{2048}}\,\,  \,
 -{\frac {231\,{t}^{5}}{8192}} \, \,
 -\left({{1463} \over {65536}} \,  +{{49} \over {1048576 }}
 \cdot \,  \lambda^2\right)
 \cdot \, t^6
\nonumber \\
  &&\hspace{-.98in} \quad  \, 
\, -\left({{4807} \over {262144}} \, +{{491} \over {4194304}}
\cdot \,  \lambda^2\right)
\cdot \, t^7
\,\,  -\left({{129789} \over {8388608}} \, +{{205491} \over {1073741824}}
\cdot \,  \lambda^2\right)
\cdot \, t^8
\,\, \, \, + \, \, \cdots
\nonumber
\end{eqnarray}
The $\, \lambda$-extension of (\ref{factor2int25first}) can {\em also}
be seen as a $\, \mu$-deformation of
the correlation function $\, C(2,5)$,  given by the exact expression 
(\ref{factor2int25first}) with (\ref{factor2int25F1}) and (\ref{factor2int25F2}), 
as a polynomial expression in $\,  {\tilde E}$  and  $\,  {\tilde K}$:
\begin{eqnarray}
\label{factor2int25M}
&& \hspace{-.9in}   \, \,\, \quad \quad \quad
C(2, \,  5; \, \lambda) \,\, \, = \, \,\,\, \,
1 \,\, \, -{{t} \over {4}} \, \, \, -{{3} \over {32}} \, t^2 \, \,\,
- {{7} \over {128}} \,t^3 \, \, \, - {{77} \over {2048}} \,t^4
   \, \,\, -{{231} \over {8192}} \, t^5
   \nonumber \\ 
&& \hspace{-.9in}   \quad \quad \quad \quad \quad  \quad
 -\Bigl( {{23457} \over {1048576}}   -{{49} \over {1048576}} \, \mu \Bigr) \cdot \, t^6 \, \, 
  -\Bigl( {{7403} \over {4194304}}   -{{491} \over {4194304}} \, \mu \Bigr) \cdot \, t^7
 \nonumber  \\ 
&& \hspace{-.9in}
 \quad   \quad  \quad   \quad \quad  \quad \quad \quad  \quad
 -\Bigl( {{16818483} \over {1073741824}} -{{205491 } \over { 1073741824}} \, \mu\Bigr)
 \cdot \, t^8
    \\
  && \hspace{-.9in}
 \quad   \quad   \quad   \quad  \quad  \quad   \quad \quad \quad \quad \, \quad
 -\Bigl( {{58337917} \over {4294967296}} \, -{{1115389} \over {4294967296}} \, \mu \Bigr)
 \cdot \, t^9
     \, \, \, \, \,\, \, + \, \, \, \cdots
     \nonumber 
\end{eqnarray}
These two series can be seen to identify if one has the following
relation between $\, \lambda$ and $\, \mu$:
\begin{eqnarray}
\label{lambdamubis}
\hspace{-.58in} \quad \quad \quad \quad
  \lambda^2 \, \, = \, \,  \,  1 \, - \, \mu
  \quad \quad \quad\quad \hbox{or:}    \quad  \quad \quad \quad
  \mu  \, \, = \, \,  \,  1 \, - \,\lambda^2. 
\end{eqnarray}

The $\, \alpha$-extension of (\ref{factor2int25F1}) reads
\begin{eqnarray}
\label{factor2int25F1alpha}
  && \hspace{-.9in}  \quad  \quad     \, \,\,
F_1(2,5; \, \alpha) \, \, = \, \, \, \,  - {{45} \over {16}} \, t^3
\, \, \,  -{{135} \over {128}} \,t^4 \,\, \, 
 -{{1485} \over {2048}} \, t^5 \,\, \, 
 -\Bigl( {{4545} \over {8192}} +{{315} \over {8192}} \, \alpha \Bigr)
 \cdot \, t^6  \,
 \nonumber  \\ 
  && \hspace{-.9in}
 \quad   \quad   \quad   \quad   \quad \quad  
 -\Bigl( {{58995} \over {131072}}  +{{17955} \over {262144}} \, \alpha \Bigr) \cdot \, t^7
 \, \,  \, \, -\Bigl( {{794745} \over {2097152}} +{{188055 } \over {2097152}} \, \alpha \Bigr)
 \cdot \, t^8
 \nonumber  \\ 
&& \hspace{-.9in}
 \quad   \quad  \quad   \quad   \quad   \quad  \quad \quad   \quad     \quad  \,
 -\Bigl( {{21971565} \over {67108864}}  +{{876645} \over {8388608}} \, \alpha \Bigr)
 \cdot \, t^9
 \, \, \, \,\,  + \, \, \, \cdots 
\end{eqnarray}
and the $\, \alpha$-extension of (\ref{factor2int25F2}) reads:
\begin{eqnarray}
\label{factor2int25F2alpha}
  && \hspace{-.9in}  \quad  \quad     \, \,\,
F_2(2,5; \, \alpha) \, \, = \, \, \,  \,   - {{45} \over {16}} \, t^3 \,  \,
+{{45} \over {128}} \,t^4 \, \, \,
     +{{315} \over {2048}} \, t^5 \, \,
\, +\Bigl( {{315} \over {4096}} +{{315} \over {8192}} \, \alpha \Bigr) \cdot \, t^6  \,
\nonumber  \\ 
  && \hspace{-.9in}
 \quad   \quad   \quad  \quad   \quad   \quad 
+\Bigl( {{11655} \over {262144}}  +{{12915} \over {262144}} \,  \alpha \Bigr) \cdot \, t^7
 \, \, +\Bigl( {{14805} \over { 524288}} +{{ 106155 } \over {2097152}} \,  \alpha \ \Bigr)
\cdot \, t^8
\nonumber  \\ 
  && \hspace{-.9in}
 \quad   \quad  \quad   \quad  \quad  \quad   \quad   \quad  \quad  \,
 +\Bigl( {{1285515} \over {67108864}}  +{{ 408555} \over {8388608}} \,  \alpha\Bigr)
\cdot \, t^9
 \,\, \,\,  + \, \, \, \cdots 
\end{eqnarray}
 One thus verifies that relation (\ref{factor2int25first}) can be ``lambda-extended''  
\begin{eqnarray}
\label{factor2int25lambda}
&& \hspace{-.9in} \quad \quad  \quad \, \,\,
C(2, \,  5; \, \lambda) \,\, \, = \, \,\,\, 
{{ 256} \over { 2025 }} \cdot \, {{(1 \, -t)^{1/2} } \over { t^6 }} \cdot \,
F_1(2,5; \, \alpha) \cdot \, F_2(2,5; \, \alpha),
\end{eqnarray}
provided:
\begin{eqnarray}
\label{relatalphaLambdabis}
&&\hspace{-.58in} \quad   \quad \,\, \,
   \mu \,\, \, = \, \, \,\, \,  4 \cdot \, \alpha \cdot \, (1 \, -\alpha)
   \quad   \quad \quad \hbox{or:}    \quad   \quad  \quad
   \lambda^2 \, \, = \, \,  \, \Bigl( 2 \, \alpha \, -1  \Bigr)^2. 
\end{eqnarray}
Again one verifies the remarkable identities:
\begin{eqnarray}
\label{relatalphaLambdabisidenti}
&&\hspace{-.9in}  \,\, \,\quad \quad  \quad \quad \quad 
F_2(2,5; \, \alpha) \, \, \, = \, \,\,
(1 \, -t)^{1/2} \cdot \, F_1(2,5; \, 1 \, -\alpha),
   \nonumber  \\ 
  && \hspace{-.9in} \quad \quad \quad \quad  \quad \,\, \,
   F_2(2,5; \, 1 \, -\alpha) \, \, \, = \, \,\,
   (1 \, -t)^{1/2} \cdot \, F_1(2,5; \, \alpha).
\end{eqnarray}
In particular one has:
\begin{eqnarray}
\label{factor2int25F1particular}
&& \hspace{-.9in}
   \, \,\,
F_1(2,5; \, 0) \, \, = \,\, \, \,
2 \cdot \,(1 \, -t) \cdot \, (2\,t \, +1) \cdot \, {\tilde K}^2 \,\, \,
+(7\,t^2 \, -15\,t \, -4) \cdot \, {\tilde E} \, {\tilde K}  \, \,
\nonumber \\ 
&& \hspace{-.9in}
 \quad \quad  \quad  \quad \quad \quad \, \,\,
+(2\,t^2 \, +13\, t \, +2) \cdot \, {\tilde E}^2, 
\end{eqnarray}
\begin{eqnarray}
\label{factor2int25F1particular2}
&& \hspace{-.9in}
  \, \,\,
F_1 (2,5; \, 1) \, \, = \, \, \,
 (1 \, -t)^{-1/2} \cdot \, \Bigl(5 \cdot \, (t-1)^3 \cdot \,  {\tilde K}^3
\, \, -(11\,t \, -17) \cdot \, (t \, -1)^2 \cdot \,  {\tilde E}\, {\tilde K}^2
\nonumber \\ 
&& \hspace{-.9in}
 \quad \quad     \quad \,\, \,
+(t \, -1) \cdot \,(2\,t^2 \, -33\, t \, +19) \cdot \, {\tilde E}^2 \,  {\tilde K}
\, \,\, +(7\, t^2 \, -22\, t \, +7) \cdot \,  {\tilde E}^3 \Bigr),
\end{eqnarray}
\begin{eqnarray}
\label{factor2int25F2particular3}
&& \hspace{-.9in}
\, \,\,
F_2(2,5; \, 0) \, \,\, = \, \, \, \,  5 \cdot \, (t-1)^3 \cdot \,  {\tilde K}^3
\, \, \, -(11\,t \, -17) \cdot \, (t \, -1)^2 \cdot \,  {\tilde E}\, {\tilde K}^2
\nonumber \\ 
&& \hspace{-.9in}
 \quad \quad \quad    \quad \quad \, \,
+(t \, -1) \cdot \,(2\,t^2 \, -33\, t \, +19) \cdot \, {\tilde E}^2 \,  {\tilde K}
\, \,\, +(7\, t^2 \, -22\, t \, +7) \cdot \,  {\tilde E}^3. 
\end{eqnarray}
\begin{eqnarray}
\label{factor2int25F1particular4}
&& \hspace{-.9in}
\, \,\,
F_2(2,5; \, 1) \, \, = \, \, \,
(1 \, -t)^{1/2} \cdot \, \Bigl(2 \cdot \,(1 \, -t)
\cdot \, (2\,t \, +1) \cdot \, {\tilde K}^2 \, \,
+(7\,t^2 \, -15\,t \, -4) \cdot \, {\tilde E} \, {\tilde K}  \, \,
\nonumber \\ 
&& \hspace{-.9in}
 \quad \quad \quad \quad  \quad  \quad \quad \quad \, \,\,
+(2\,t^2 \, +13\, t \, +2) \cdot \, {\tilde E}^2\Bigr). 
\end{eqnarray}
The series expansions of the previous exact expressions read:
\begin{eqnarray}
\label{F1F2a0}
 && \hspace{-.9in}  \, \,\, 
F_1\Bigl(2,5; \, 0\Bigr) \, \, = \, \,\,
-{{45 } \over { 16}} \, t^3 \,\, \,  - {{135 } \over {128}} \, t^4 \,\, 
- {{1485 } \over {2048}} \,t^5\, 
\, - {{4545 } \over {8192}} \,t^6 \, \, -{{ 58995 } \over { 131072}} \, t^7
\,  \, \,  + \, \,  \, \cdots
\nonumber \\
  && \hspace{-.9in}  \, \,\,
     F_1\Bigl(2,5; \, 1\Bigr) \, \, = \, \,\,
-{{45 } \over { 16}} \, t^3 \,\, \,- {{135 } \over {128}} \, t^4 \,\,
- {{1485 } \over {2048}} \,t^5\,
\, - {{1215 } \over {2048}} \,t^6 \,\, -{{135945  } \over { 262144}} \, t^7
\,  \,  \, + \, \,  \, \cdots
\end{eqnarray}
\begin{eqnarray}
\label{F1F2a1}
 && \hspace{-.9in}  \, \,\, 
 F_2\Bigl(2,5; \, 0\Bigr) \, \, = \, \, \,
 -{{45 } \over { 16}} \,t^3 \,\,\,  +{{45 } \over {128}} \,t^4\, 
 \, +{{315 } \over {2048}} \,t^5 \,\, 
 +{{315 } \over {4096}}  \,t^6 \,\,  +{{ 11655 } \over { 262144}} \,t^7
 \,  \, \,  + \, \,  \, \cdots 
\nonumber \\
&& \hspace{-.9in}   \, \,\,
 F_2\Bigl(2,5; \, 1\Bigr) \, \, = \, \, \,
 -{{45 } \over { 16}} \,t^3 \,\, \,+{{45 } \over {128}} \,t^4
 \,\, +{{315 } \over {2048}} \,t^5 \,\,
 +{{945 } \over { 8192}} \,t^6 \,\, +{{ 12285 } \over {131072}} \,t^7
 \,  \,  \, + \, \,  \, \cdots 
\end{eqnarray}
It is worth noticing that (in contrast with the $\, \lambda$-extension
$\, C(2,5; \,\lambda)$), the $\, F_n(2,5; \, \alpha)$'s have {\em two}
different values of the parameter $\, \alpha$ for which these
$\, \alpha$-extensions are D-finite, being (homogeneous)
polynomials in $\, {\tilde E}$ and $\, {\tilde K}$. One remarks with
that the corresponding  polynomials in $\, {\tilde E}$ and $\, {\tilde K}$
{\em are not necessarily of the same degree}
in  $\, {\tilde E}$ and $\, {\tilde K}$.

\vskip .1cm
{\bf Remark: the $\, \alpha \, = \, 1/2$ algebraic subcase.} 
For $\, \alpha \, = \, 1/2$ the corresponding $\, \lambda$ deduced
from (\ref{relatalphaLambdabis})  is $\, \lambda \, = \, \, 0$
and the two product formula (\ref{factor2int25lambda}) becomes
\begin{eqnarray}
\label{factor2int25lambdaa1/2}
&& \hspace{-.9in}  
C(2, \,  5; \, 0) \,\,  = \, \,\,
{{ 256} \over { 2025 }} \cdot \, {{(1 \, -t)^{1/2} } \over { t^6 }} \cdot \,
F_1\Bigl(2,5; \, {{1} \over {2}}\Bigr) \cdot \, F_2\Bigl(2,5; \, {{1} \over {2}}\Bigr)
\,\,  = \, \,\, (1 \, -t)^{1/4},
\end{eqnarray}
in agreement with the expansion (\ref{factor2int25M}) evaluated at $\, \lambda \, = \, \, 0$. 
Using the identity (\ref{relatalphaLambdabisidenti}) one gets 
\begin{eqnarray}
\label{relatalphaLambdabisidenti2}
&&\hspace{-.9in}  \,\, \,\quad \quad  \quad \quad \quad \quad 
   F_2\Bigl(2,5; \, {{1} \over {2}}\Bigr)  \, \, \, = \, \, \,
   (1 \, -t)^{1/2} \cdot \, F_1\Bigl(2,5; \, {{1} \over {2}}\Bigr),
 \end{eqnarray}
 which enables to write (\ref{factor2int25lambdaa1/2}) as:
\begin{eqnarray}
\label{factor2int25lambdaa1/2bis}
&& \hspace{-.9in}  \quad \quad  \quad \quad \quad 
C(2, \,  5; \, 0) \,\,  = \, \,\,
{{ 256} \over { 2025 }} \cdot \, {{1 } \over { t^6 }}
\cdot \, \Bigl(F_2\Bigl(2,5; \, {{1} \over {2}}\Bigr)\Bigr)^2
\,\,  = \, \,\, (1 \, -t)^{1/4},
\end{eqnarray}
from which one deduces
\begin{eqnarray}
\label{deduces}
  && \hspace{-.9in}    \quad \quad 
F_2\Bigl(2,5; \, {{1} \over {2}}\Bigr)
\, \, \, = \, \, \, - \, {{ 45} \over { 16}} \cdot \, t^3  \cdot \, (1\, -t)^{1/8}
\nonumber \\
&& \hspace{-.9in}  \quad \quad \quad
\, \, \, = \, \, \,
-{{45 } \over { 16}} \,t^3 \,\,\,  \,+{{45 } \over {128}} \,t^4\, 
 \, +{{315 } \over {2048}} \,t^5 \,\, 
 +{{1575 } \over {16384}} \,t^6 \, \, +{{36225 } \over {524288}} \,t^7
 \,  \,\, \, \,+ \, \,  \, \cdots
\end{eqnarray}
or:
\begin{eqnarray}
\label{deduces2}
  && \hspace{-.9in}   \quad \quad 
F_1\Bigl(2,5; \, {{1} \over {2}}\Bigr)
\, \, \, = \, \, \, \, - \, {{ 45} \over { 16}} \cdot \, t^3  \cdot \, (1\, -t)^{-3/8}
\nonumber  \\
&& \hspace{-.9in}  \quad \quad \quad
\, \, \, = \, \, \,\, 
-{{45 } \over { 16}} \, t^3 \,\,\, - {{135 } \over {128}} \, t^4 \,\, 
- {{1485 } \over {2048}} \,t^5\, 
\, - {{9405 } \over {16384}} \,t^6 \,\,  -{{253935  } \over {524288}} \, t^7
\,  \,\, \,\,  + \, \,  \, \cdots
\end{eqnarray}

\vskip .2cm

\subsection{Form factor deformation around the algebraic function  $\, F_1\Bigl(2,5; \, {{1} \over {2}})$}
\label{deformF1}
Introducing a form-factor $\, \beta$-deformation around the algebraic function (\ref{deduces2})
($\beta$ is the deformation parameter around $\, \alpha=\, 1/2$)
\begin{eqnarray}
\label{deduces2beta}
  && \hspace{-.9in}   \quad \quad \quad 
F_1\Bigl(2,5; \, {{1} \over {2}}\, + \, \beta\Bigr)
\, \, \, = \, \, \,
- \, {{ 45} \over { 16}} \cdot \, t^3  \cdot \, (1\, -t)^{-3/8}
\, \, \,\,   + \beta \cdot \, G(t) \,\, \, + \, \, \, \cdots 
\end{eqnarray}
and inserting (\ref{deduces2beta}) in the non-linear ODE verified by
(\ref{deduces2beta}), one gets an {\em order-three linear differential operator}
which is the direct sum of an order-one linear differential
operator and an order-two  linear differential operator, yielding the following exact
expression for $\, G(t)$ in (\ref{deduces2beta}):
\begin{eqnarray}
\label{deduces2beta1}
&& \hspace{-.98in} 
G(t) \, \, = \, \, \, -{{45} \over {16}} \cdot \, t^3 \cdot \, (1\, -t)^{-3/8} \,\,\,\,
\, -{{9} \over {16}} \cdot \,  (1\, -t)^{-3/8}  \cdot \,
P_{E,K}  
\\
&& \hspace{-.98in} 
\, \, = \, \, \,\, 
-{{315} \over { 8192}} \cdot \, t^6 \, -{{17955} \over {262144}}  \cdot \, t^7 \,
-{{188055} \over {2097152}}  \cdot \, t^8
\, -{{876645} \over {8388608}}  \cdot \, t^9 \,
-{{1929015} \over {16777216}}  \cdot \, t^{10}
\,\, + \,\,\, \cdots
\nonumber 
\end{eqnarray}
where $\, P_{E,K} \, \, $ is a polynomial in $\, {\tilde E}$ and $\, {\tilde K}$: 
\begin{eqnarray}
\label{deduces2betawhere}
&& \hspace{-.9in} \quad \quad 
P_{E,K} \, \, = \, \, \,
4 \cdot \, t^2 \cdot \,(t-1) \cdot \, (t^2-6\,t+16)  \cdot \, {{ d {\tilde K}} \over {dt}}
\, \,\,
+  \,t^2 \cdot \,(2\,t^2 \, -13\,t \, +16) \cdot \, K
\nonumber \\
&& \hspace{-.9in} \quad \quad \quad \quad \quad 
\, \, = \, \, \,   t \cdot  \, (t^2 \, -28\,t \, +32) \cdot \, {\tilde K}
   \, \,\, \, -2 \cdot \,  (t^2 \, -6\,t \, +16) \cdot \, {\tilde E}.
\end{eqnarray}
As far as the log-derivative with respect to $\, t$ is concerned, one gets:
\begin{eqnarray}
\label{deduces2betalog}
 && \hspace{-.9in}   \quad \quad \quad \quad \quad 
t \cdot \, (t \, -1) \cdot \,
{{ d } \over {dt}} \ln \Bigl(F_1\Bigl(2,5; \, {{1} \over {2}}\, + \, \beta \Bigr)\Bigr)
\, \, \, = \, \, \,
-3 \,\,\,  + {{21} \over {8}} \cdot \, t  \,
\nonumber \\
  && \hspace{-.9in}   \quad \quad  \quad \quad \quad \quad \quad \quad \quad \quad
 \, \, \, + \, \, \beta  \cdot \, {{t \, -1 } \over { 5 \, t^3 }}
 \cdot \, \Bigl( t \cdot \,{{ d P_{E,K}} \over {dt}} \, - \, 3 \cdot \,  P_{E,K} \Bigr)
   \, \, \, \, + \, \, \, \cdots 
\end{eqnarray}
where the first deformation term is also polynomial in $\, {\tilde E}$ and $\, {\tilde K}$.

\vskip .2cm

\section{Comments and speculations on the lambda-extensions of the two-point correlation functions.}
\label{Coments}

The previous sections provide an illustration of nice involutive symmetries of $\, \alpha$-extension solutions of  Painlev\'e-like
non-linear ODEs (see (\ref{identity})). Furthermore, recalling  (\ref{particular3}),
(\ref{particular31}),   (\ref{f_3alpha1/2})  and  
(\ref{particular4}),  (\ref{particular41}),   (\ref{f_4alpha1/2}), namely
\begin{eqnarray}
\label{particular3bis}
 &&\hspace{-.98in} \,\, \,
     f_3(0)\, =\,\, \,  (t\, -2) \cdot \, {\tilde E}\,\, 
+2 \cdot \, (1\, -t) \cdot \, {\tilde K}, \nonumber \\
&&\hspace{-.98in} \,\, \, 
   \label{particular31bis}
   f_3(1)\, =\,\, \,
   (1\, -t)^{-1/4} \cdot \,
   \Bigl( 3 \, {\tilde E}^2\,\, +2 \cdot \, (t\, -2) \cdot \,{\tilde E}\, {\tilde K}
\,\,  +(1-t) \cdot \, {\tilde K}^2  \Bigr), \, \\
&&\hspace{-.98in} \,\, \, 
   \label{f_3alpha1/2bis}
 f_3\Bigl({{1} \over {2}} \Bigr)  \, \, = \,\,\, 
-{{3} \over {8}} \cdot \, t^2 \cdot \, (1\, -t)^{1/16} \cdot \,
\Bigl( {{ 1 \, +(1-t)^{1/2}} \over {2 }}   \Bigr)^{-3/4} \cdot \,
   \Bigl( 2 \cdot \, {{ ( 1\, -(1\, -t)^{1/2}) } \over {t}}  \Bigr)^{1/2},
   \nonumber
\end{eqnarray}
and
\begin{eqnarray}
\label{particular4bis}
  &&\hspace{-.98in} \,\, \,
     f_4(0)\, = \,\,\,\, 
3 \, {\tilde E}^2\,\,\, +2 \cdot \, (t\, -2) \cdot \,{\tilde E}\, {\tilde K}
     \,\, \, +(1-t) \cdot \, {\tilde K}^2,
\nonumber \\
&&\hspace{-.98in} \,\, \,  
 \label{particular41bis}
 f_4(1)\, =\,\, \,  \,   (1\, -t)^{1/4} \cdot \,  \Bigl( (t\, -2) \cdot \, {\tilde E}\,\, 
   +2 \cdot \, (1\, -t) \cdot \, {\tilde K}  \Bigr),\, \\
&&\hspace{-.98in} \,\, \,  
 \label{f_4alpha1/2bis}
   f_4\Bigl({{1} \over {2}} \Bigr) \,\, = \,\, \,\, \,
-{{3} \over {8}} \cdot \, t^2 \cdot \, (1\, -t)^{1/16} \cdot \,(1\, -t)^{-1/4} \cdot \,
   \Bigl( {{ 1 \, +(1-t)^{1/2}} \over {2 }}   \Bigr)^{-3/4}
   \nonumber \\
&&\hspace{-.98in} \quad \quad \quad   \quad \quad \quad \quad \quad \quad \quad
   \times \,   \Bigl( 2 \cdot \, {{ ( 1\, -(1\, -t)^{1/2}) } \over {t}}  \Bigr)^{1/2},
   \nonumber
\end{eqnarray}
we see that the $\, \alpha$-extension $\, f_3( \alpha)$  (resp. $\, f_4( \alpha)$)  has three
different values of the parameter $\, \alpha \, $ for which the corresponding 
$\, \alpha$-extensions are D-finite being (homogeneous) polynomials in $\, {\tilde E}$ and $\, {\tilde K}$
of {\em different degree} in $\, {\tilde E}$ and $\, {\tilde K}$. It is straightforward to see that
$\, f_3( \alpha)$  (resp. $\, f_4( \alpha)$) is {\em not}  a linear interpolation of these three
D-finite expressions. For generic values of $\, \alpha$,  $\, f_3( \alpha)$  (resp. $\, f_4( \alpha)$)
is {\em not}  D-finite\footnote[1]{In section 4.1 of~\cite{bm} we provide, not a proof, but  arguments strongly suggesting that
  such lambda-extensions are not generically D-finite.},
it is differentially algebraic~\cite{Tutte,Moore,BoshernitzanRubel}, being  
solution of a Painlev\'e-like non-linear ODE. Let us now display several remarkable properties
of such lambda-extensions.

\subsection{Other remarkable features of the lambda-extensions of the two-point correlation functions.}
\label{subComents}

In fact $\, \alpha = \, 1/2 \, $ is not the only value of $\, \alpha$ for which $\, f_3( \alpha)$  (resp. $\, f_4( \alpha)$)
becomes an algebraic function. One has an {\em infinite number} of (algebraic) values of
$\, \alpha$ for which $\, f_3( \alpha)$  (resp. $\, f_4( \alpha)$) becomes an algebraic function. 
This phenomenon is illustrated in detail in~\cite{bm} in the case of the lambda-extension
of the diagonal\footnote[5]{Recall that diagonal correlation functions depend only on $\, k \, = \sqrt{t}$. They are
independent of the anisotropic parameter $\, \nu$.} correlation function $\, C(1,\, 1)$, but one has similar results for other
non-diagonal two-point correlation functions (at $\, \nu \, = \, -k$),
or for factors of the correlation functions like  the $\, f_i( \alpha)$'s.

For pedagogical reasons we restrict our analysis to the 
low-temperature two-point correlation function
$\, C(1,1)$ and its lambda extension.
For instance, the form factor expansion of the lambda extension
of this low-temperature correlation function reads
\begin{eqnarray}
\label{form11fact}
\hspace{-0.98in}&& \quad \quad  \quad \quad   \quad \quad
C_{-}(1, \, 1; \, \lambda) \, \, = \, \, \, \,
(1\, -t)^{1/4} \cdot \,
\Bigl( 1 \, + \, \sum_{n=1}^{\infty} \, \lambda^{2\, n} \cdot \, f_{1, \, 1}^{(2\, n)} \Bigr), 
\end{eqnarray}
where the first form factors read:
\begin{eqnarray}  
  \label{form11f11}
\hspace{-0.98in}&& \quad \quad \quad \quad 
f_{1, \, 1}^{(2)} \, \, = \, \, \,  {{1} \over {2}} \cdot \,
\Bigl(1 \,\,\, -3 \, E \, K \,\, -(t\, -2) \cdot \, K^2\Bigr), 
  \\
  \hspace{-0.98in}&& \quad \quad \quad \quad
 \label{form11f114}
f_{1, \, 1}^{(4)} \, \, = \, \, \,
{{1} \over {24}} \cdot \,
\Bigl(9 \,\, \, -30\, {\tilde E} \, {\tilde K}\,\, -10 \cdot \, (t\,-2) \cdot \, {\tilde K}^2
 \nonumber  \\
  \hspace{-0.98in}&&   \quad \quad \quad \quad \quad  \quad  \quad 
  \, +(t^2\, -6t \, +6) \cdot \, {\tilde K}^4 \,\, +15 \, {\tilde E}^2\, {\tilde K}^2 \, \,
     +10\cdot (t\, -2) \cdot \,  {\tilde E}\, {\tilde K}^3 \Bigr).
\end{eqnarray}
For $\, \lambda \, = \, 1$ we must recover, from (\ref{form11fact}),
the well-known expression of
the {\em low-temperature} two-point correlation function $\, C(1,1) \, = \, {\tilde E}$:
\begin{eqnarray}
\label{form11E}
\hspace{-0.98in}&& \quad   \, \quad  \quad  
C_{-}(1, \, 1; \, 1) \, \, = \, \, \,  E \, \,  = \, \, \,\,
1 \,\, -{{1} \over {4}} \cdot \, t   \,\,  \, -{{3} \over {64}} \cdot \, t^2 \,\,
- {{5} \over {256}} \cdot \, t^3 \,
  \,  -{{175} \over {16384}} \cdot \, t^4 \, \,  \,  \, + \, \, \, \cdots 
\nonumber   \\
 \hspace{-0.98in}&& \quad \quad \quad \quad  \quad  \quad \quad \quad 
 \, = \, \, \,\,
 (1\, -t)^{1/4} \cdot \,
\Bigl( 1 \, \, + \, \sum_{n=1}^{\infty}  \, f_{1, \, 1}^{(2\, n)} \Bigr), 
\end{eqnarray}
which corresponds to write the ratio $\, {\tilde E}/(1\, -t)^{1/4} \, $ as an infinite sum
of polynomial expressions of $\, {\tilde E}$ and $\, {\tilde K}$, thus yielding
a non-trivial {\em infinite sum identity} on the complete elliptic integrals  $\, {\tilde E}$ and $\, {\tilde K}$.

Since all these lambda extensions are power series in $\, t$, we can also try
to get, order by order, the  series
expansion of   $\, C_{-}(1, \, 1; \, \lambda) \, $
from the corresponding non-linear ODE  (see (\ref{jmequation}) below). Recalling~\cite{Holonomy} the
form factor expansion (\ref{form11fact}),
we can either see the series expansion in $\, t$ as a deformation of the simple
algebraic function $\, (1 \, -t)^{1/4}$, or {\em more naturally}, see
the series expansion of the lambda-extension
of the low-temperature two-point correlation function $\, C_{-}(1, \, 1; \, \lambda) \, $
as a deformation of
the exact expression $\, C_{-}(1,1) \, = \, {\tilde E}$ (here $M$ denotes here a difference
to $\, \lambda^2= \, 1$, namely $\, M \, = \, 4 \cdot \, (1\, -\lambda^2)$):
\begin{eqnarray}
\label{morenaturallyM1}
\hspace{-0.98in}&& \, \, \, \quad   \quad   \quad  
C_{-}(1, \, 1; \, \lambda) \,   \, = \, \, \,  C_{M}(1, \, 1; \, M)
\nonumber \\
  \hspace{-0.98in}&& \, \, \,
\quad   \quad   \quad  \quad   \quad   \quad   \quad    \, \, = \, \, \,\,\,
{\tilde E}\,\, \,\,  + M \cdot g_1(t)\, \,\, + M^2 \cdot \, g_2(t) \,
\,\, + M^3 \cdot \, g_3(t) \,  \,  \,  \,\, + \, \, \cdots 
\end{eqnarray}

All the $\, g_n(t)$'s in (\ref{morenaturallyM1}) are also~\cite{bm} polynomials\footnote[9]{This cannot be deduced
  straightforwardly from an identification of two representations (\ref{morenaturally}) and (\ref{actually2}) 
  of the lambda extension $\, C_{-}(1, \, 1; \, \lambda)$. This identification yields an infinite number of (infinite sum)
  non-trivial identities on  $\, {\tilde E}$ and $\, {\tilde K}$.} in  $\, {\tilde E}$ and $\, {\tilde K}$. For
instance $\, g_1(t)$ in  (\ref{morenaturally}) reads: 
\begin{eqnarray}
\label{gg1}
\hspace{-0.98in}&&
\quad   \quad \quad \quad\quad  \quad
g_1(t) \, = \, \,\, \,  {{1} \over {24}} \cdot \, {\tilde E} \, \,\,\,
- \,   {{1} \over {8}} \cdot \, {\tilde K} \, {\tilde E}^2 \, \,\,
   - \,   {{ t \, -1 } \over {12}} \cdot \,  {\tilde K}^3.
\end{eqnarray}

Using the sigma-form of Painlev\'e VI equation (\ref{jmequation})
one can find that this expansion (\ref{morenaturallyM1})
  reads as a series expansion in the variable $\, t$:
\begin{eqnarray}
\label{morenaturally}
\hspace{-0.98in}&& \, \, \,  \, \, \, 
 C_{M}(1, \, 1; \, M) \, \, = \,\, \, \,\, 
 1 \, \, \,\, -{{1} \over {4}} \cdot \, t \,\, \,  \,
 \, -\Bigl({{3} \over {64}}  \, +{{3} \over {256}}\cdot \, M \Bigr) \cdot \, t^2
 \, \,\, \,
 -\Bigl( {{5} \over {256}} \, + {{9} \over {1024}}  \cdot \, M    \Bigr)   \cdot \, t^3
\nonumber   \\
 \hspace{-0.98in}&& \quad \quad \quad \quad
 \, \, -\Bigl(  {{175} \over {16384}} \,   + {{441} \over {65536}}  \cdot \, M     \Bigr)
 \cdot \, t^4\,  \,  \,  \,
 -\Bigl(  {{441} \over {65536}} \,   +{{1407} \over {262144}} \cdot \,  M   \Bigr)
 \cdot \,t^5
 \nonumber   \\
 \hspace{-0.98in}&& \quad \quad  \quad \quad \quad \quad  \, \,
 -\Bigl({{4851} \over {1048576}}  \,  \,
 +{{9281} \over {  2097152}}    \cdot \, M  \,  -{{5} \over {  16777216 }}  \cdot \, M^2 \Bigr)
 \cdot \, t^6
    \,\, \, \,  \, \, + \, \, \cdots 
\end{eqnarray}

\vskip .1cm 
\subsubsection{Deformation around an algebraic subcase. \\}
\label{subsub}
Recalling that
 one finds~\cite{bm} that (\ref{morenaturally}) is actually, for $\, M\, = \, 2$, the series expansion 
 of an {\em algebraic function} (see (\ref{actually2G0}) below), one can try to write
 the series (\ref{morenaturally}) as a deformation of this
 $\, M\, = \, 2 \, $ algebraic function (\ref{actually2G0})
\begin{eqnarray}
\label{actually2}
\hspace{-0.98in}&& \, \, \quad  \quad  \quad   \quad  \quad   \quad 
C_{\rho}(1, 1; \rho) \,  \, = \, \,\,  \, \, G_0(t)
\, \, \, \,  \, +   \rho \cdot \, G_1(t) \,\,\, \, +  \rho^2 \cdot \, G_2(t)
\, \,\, \,\,  + \, \,\, \,  \cdots
\end{eqnarray}
where
\begin{eqnarray}
\label{actually2G0}
\hspace{-0.98in}&& \, \,  \quad   \quad \quad   \quad \quad   \quad \quad  
G_0(t) \,  = \, \,  \,  \,
 (1-t)^{1/16} \cdot \, \Bigl( {{1 \, \, +(1-t)^{1/2}} \over {2}} \Bigr)^{3/4}, 
\end{eqnarray}
and where $\, \rho \, = \, M \, -2$.
Again one can ask whether
the $\, G_n(t)$'s in (\ref{actually2}) are D-finite, and, again, polynomials
in the complete elliptic integrals $\, {\tilde E}$ and $\, {\tilde K}$.
This is actually the case. One  finds that
(\ref{actually2}) can be written as
\begin{eqnarray}
\label{actually2rho}
\hspace{-0.98in}&& \quad  
{{C_{\rho}(1, 1; \rho)} \over {G_0(t) }}  \,  \, \,  = \, \,\,  \,   1
\, \, \, \, +  \rho \cdot \, \Bigl({{1} \over {4}} \cdot \, S_2  \, \,  -{{1} \over {4}} \Bigr)
\,\,\,\,  + \rho^2 \cdot \, 
\Bigl( {{1} \over {32}} \cdot \, S_3
\,  \,- \, {{1} \over {16}} \cdot \, S_2   \,\, + {{3} \over {32}}   \Bigr)
\nonumber \\
 \hspace{-0.98in}&&  \quad \quad   \quad    \quad  
 +  \rho^3 \cdot \,
 \Bigl(   {{1} \over {384}} \cdot \, S_4  \, \, -  {{1} \over {128}} \cdot  \, S_3 \, \,
                    +  {{13} \over {384}} \cdot  \, S_2  \,  \, - {{5} \over {128}}  \Bigr)
\, \, \, \, + \, \, \, \cdots 
\end{eqnarray}
where:
\begin{eqnarray}
\label{actually2rho}
\hspace{-0.98in}&& 
S_2 \, \, = \, \, \, {{ 2} \over { t }} \cdot \,
\Bigl(1  \,\, -(1\, -t)^{1/2}\Bigr) \cdot \, {\tilde E}
\, \, \,  \,  - {{ 1} \over { 2\, t }} \cdot \,
\Bigl( (t\, -4) \cdot \, (1\, -t)^{1/2} \,  \, -(3\, t \, -4) \Bigr) \cdot \, {\tilde K},
\nonumber \\  
\hspace{-0.98in}&&
S_3 \, \, = \, \, \, {{1} \over { 4}} \cdot \,
\Bigl( 6 \cdot \, (1\, -t)^{1/2} \, \,  -(t \, -2) \Bigr) \cdot \, {\tilde K}^2 \, \,  \, \, -3 \, {\tilde E}\, {\tilde K}, 
\nonumber
\end{eqnarray}
\begin{eqnarray}
\hspace{-0.98in}&&
S_4 \, \, = \, \, \,
{{3} \over { t}} \cdot \, \Bigl( (t\, -4) \cdot \, (1\, -t)^{1/2} \,\,  \, -(3\, t \, -4) \Bigr)
\cdot \, {\tilde E} \, {\tilde K}^2 \,\, \,
-{{ 6} \over { t }} \cdot \, (1 \, \, -(1\, -t)^{1/2}) \cdot \, {\tilde E}^2 \, {\tilde K}
\nonumber \\  
\hspace{-0.98in}&& \quad \quad  \quad \quad 
\, + \, {{1} \over { 8 \, t}} \cdot \, \Bigl( (t^2 \, -28\,t \, +48) \cdot \, (1\, -t)^{1/2}
\,  \, -(21\, t^2 \, -68\, t \, +48) \Bigr) \cdot \, {\tilde K}^3,  \,
\nonumber
\end{eqnarray}
             
\vskip .2cm

We thus see the same phenomenon as the one sketched in section (\ref{formfactorf1}) for the
$\, \alpha$-extension $\, f_1(\alpha)$ and section (\ref{deformF1})
for the $\, \alpha$-extension $\, F_1(2, 5; \, \alpha)$, seen as deformations of
algebraic function subcases. 

\vskip .2cm 

{\bf Remark:}
All these  $\, g_n(t)$'s  or $\, G_n(t)$'s are
{\em globally bounded series}\footnote[1]{A series
with rational coefficients and non-zero radius of convergence
is  a globally bounded series~\cite{ChristolDiag} if it can be recast into a series with integer
coefficients with one rescaling $\,\, t \, \rightarrow \, \, N \, t \,$
where $\, N$ is an integer.}~\cite{ChristolDiag}. This is a consequence of the fact that they are
polynomial expressions in $\, {\tilde E}$ and $\, {\tilde K}$: they are not only
D-finite, they can actually be seen to be {\em diagonals of rational functions}~\cite{ChristolDiag}. 
We have actually seen, so many times in physics, and in particular in the two-dimensional
Ising model,  the emergence of globally bounded series as
a consequence of the frequent occurrence of
{\em diagonals of rational functions}~\cite{ChristolDiag,Heun} (or $\, n$-fold
integrals~\cite{Khi5}). 
In contrast the  lambda extension $\, C_{-}(1, \, 1; \, \lambda)$ which is an infinite sum of globally bounded series
is, at first sight, a differentially algebraic function which has no reason to correspond to a globally bounded series.
     
\vskip .2cm 

\subsection{Arithmetic properties of the lambda-extensions and globally bounded series.}
\label{globally}
  
Let us  consider the series expansion (\ref{morenaturally})
for values of the parameter $\, M \, \ne 0$
not yielding the previous algebraic function series
(i.e. $\, M \, \ne \ 4 \cdot \, \sin^2(\pi m/n)$ where $\, m$ and $\, n$ are integers).

Let us change $\, t \, $ into $\,\, 16 \, t \,$ in the series expansion
(\ref{morenaturally}). One gets the following expansion:
\begin{eqnarray}
\label{glob}
\hspace{-0.98in}&& \,  \, \, \, \, 
1 \, \, \, -4\, t \, \, -(12+3\, M)  \cdot \, t^2 \, \, \, -(80+36\, M) \cdot \, t^3
\,\, \,  -(700+441\, M) \cdot \, t^4
\nonumber \\
  \hspace{-0.98in}&& \quad
\, -(7056+5628\, M) \cdot \, t^5
\, \, \,   -(77616 +74248\, M -5\, M^2) \cdot \, t^6
\nonumber \\
  \hspace{-0.98in}&& \quad\,
  -(906048 +1004960\, M -220\, M^2) \cdot \, t^7 \,
  -(11042460 +13877397\, M-6255\, M^2) \cdot \, t^8
 \nonumber \\
 \hspace{-0.98in}&& \quad \,
 -(139053200 +194712812\, M-146500\, M^2) \cdot \, t^9
\nonumber \\
\hspace{-0.98in}&& \quad 
\, -(1796567344 +2767635832\, M  -3079025\, M^2) \cdot \, t^{10}
\, \, \, \, \, + \, \,\, \cdots 
\end{eqnarray}
For integer values of $\, M$ one sees, very clearly, that the series (\ref{glob})
becomes a {\em differentially algebraic\footnote[2]{They are solutions of a non-linear ODE,
  the sigma-form of Painlev\'e VI.} series with integer coefficients}. One thus has
a first example of an {\em infinite number of differentially algebraic series with integer coefficients}.
As far as integer values of $\, M$ are concerned we have seen~\cite{bm} that the  lambda extension $\, C_{-}(1, \, 1; \, \lambda)$
is a simple algebraic function for $\, M= \, \, 2,  \, 4$ and slightly more involved  algebraic functions
for $\, M= \,  \, 1, \, 3$, and corresponds to  $\, {\tilde E} \, \, $ for $\, M= \, 0$.
These series (\ref {glob}) are, at first sight, {\em differentially algebraic}~\cite{Tutte}: is
it possible that such series could become D-finite for selected values integer of $\, M$
different from $\, M=0, \, 1, \, 2, \, 3, \, 4$ ? 

In section (4.1) of~\cite{bm} we give some strong argument to discard, at least for $\, M= \, 5$,  the possibility that
the series expansion (\ref{morenaturally}) (or the
series expansion (\ref{glob})) could be D-finite. It is differentially algebraic.

More generally,  one can see that the series expansion (\ref{morenaturally}) (or the
series expansion (\ref{glob})) is a  {\em globally bounded series}
 when $\, M$
is {\em any rational number}. 
One thus generalizes the quite puzzling result that an {\em infinite number of}
(at first sight ...) {\em differentially algebraic series} can be
{\em globally bounded series}.

\vskip .1cm
{\bf Remark:}
Quite often we see the emergence of {\em globally bounded series}~\cite{ChristolDiag} 
as solutions of  D-finite
linear differential operators, and more specifically as
{\em diagonals of rational functions}~\cite{ChristolDiag,Heun}
(this is related to the so-called Christol's conjecture~\cite{Christolconj}).
The emergence of {\em globally bounded series} that are not D-finite 
(not diagonals of rational functions) is more puzzling.
It can  be tempting to imagine that such 
{\em differentially algebraic globally bounded} situations could correspond to particular ratios of D-finite
functions\footnote[5]{Let us recall that {\em ratios} of D-finite expressions are {\em not} (generically)
  D-finite: they are  {\em differentially algebraic}~\cite{Tutte}.}, namely
ratios of diagonals of rational functions (or even rational functions
of diagonals), or even composition of diagonal of rational functions. Our prejudice is that this
is not the case, but discarding these simple scenarii is extremely difficult.

\vskip .2cm  

\section{More non-linear ODEs of the Painlev\'e type and more  $\lambda$-extensions.}
\label{Mangazeev}

In~\cite{Mangazeev} V.V. Mangazeev and A. J. Guttmann derived the following Toda-type recurrence relation
for the $\lambda$-extension $\, C(N, \, N; \, \lambda)$ of the diagonal correlation functions of the square Ising model
(see equation (6) in~\cite{Mangazeev}):
\begin{eqnarray}
\label{Toda}
  \hspace{-0.98in}&& \,  \, \, \, \,\, \,
t \cdot \, {{d^2} \over {dt^2}}  \ln(C_N)  \, +  {{d} \over {dt}}  \ln(C_N)
   \, \, \, +{{N^2} \over {1 \, -t}^2} \, \,  \,\, = \, \, \,  \,\,
   {{N^2 \, -1/4} \over {1 \, -t}^2} \cdot \, {{ C_{N-1} \cdot \,  C_{N+1} } \over {C_N^2 }}, 
\end{eqnarray}
where $\, C_N$ denotes the  $\lambda$-extensions of the low (resp. high) diagonal correlation functions
$\, C_N \, = \, \, C(N, \, N)$.  Introducing the ratio
\begin{eqnarray}
\label{Toda2}
  \hspace{-0.98in}&& \,  \, \, \, \,\, \, R_N \, \, = \, \,\, {{ C_{N-1} \cdot \,  C_{N+1} } \over {C_N^2 }}
                     \quad  \quad \hbox{or:} \quad \quad \quad
        P_N \, \, = \, \,\,  {{N^2 \, -1/4} \over {1 \, -t}^2} \cdot \, {{ C_{N-1} \cdot \,  C_{N+1} } \over {C_N^2 }},                 
\end{eqnarray}
one can easily deduce from (\ref{Toda}) (together with the same relation (\ref{Toda}) where $\, N$ is changed into $\, N-1$ and  $\, N+1$) 
other relations like:
\begin{eqnarray}
\label{Toda3}
  \hspace{-0.98in}&& \,
   t \cdot \, {{d} \over {dt}} \Bigl(  t \cdot \, {{d \ln(R_N)} \over {dt}}    \Bigr)   \, \, \, \, \, + {{2} \over{ (1\, -t)^2}}
                  \\
  \hspace{-0.98in}&& \, \quad 
 \, \, = \, \,  \, {{(N \, -1)^2 \, -1/4} \over {1 \, -t}^2} \cdot \,   R_{N-1} \, +  {{(N \, +1)^2 \, -1/4} \over {1 \, -t}^2} \cdot \,   R_{N+1} \,\,
  -2 \cdot \, {{N ^2 \, -1/4} \over {1 \, -t}^2} \cdot \,   R_{N},
 \nonumber        
\end{eqnarray}
or:
\begin{eqnarray}
\label{Toda3b}
  \hspace{-0.98in}&& \, \quad \quad \quad  \quad  \quad 
 \Bigl( t \cdot \, {{d} \over {dt}}  \Bigr)^2   \,  \ln( P_N)
    \,  \,  \, \, + {{2} \over{ 1\, -t}} \, \, \, \, = \, \,  \,  \,   \,  P_{N-1} \, +  P_{N+1} \,  -2 \, P_{N},   
\end{eqnarray}
Let us now consider, for instance, the low-temperature $\,T< \,T_c$ diagonal correlation functions. One knows that they verify
the sigma-form of Painlev\'e VI equation
\begin{eqnarray}
\label{jmequation}
\hspace{-0.98in}&& \quad  
\left(t \cdot \, (t-1) \cdot \, \frac{d^2\sigma}{dt^2}\right)^2
\\
\hspace{-0.98in}&&  \quad  \,    \quad  
\, = \, \,\, N^2 \cdot \,
\left((t -1) \cdot  \, \frac{d\sigma}{dt} \, -\sigma\right)^2
\, -4 \cdot \,\frac{d\sigma}{dt} \cdot \,
\left((t -1) \cdot \frac{d\sigma}{dt}\, -\sigma\, -\frac{1}{4}\right)
\cdot \,\left(t\frac{d\sigma}{dt}-\sigma\right).
\nonumber 
\end{eqnarray}
with
\begin{equation}
\label{sigmam}
\hspace{-0.4in}
\sigma\, \,= \,\,\,
t \cdot \, (t-1) \cdot \, \frac{d}{dt}\ln C(N,N) \, \, \, \, -\frac{t}{4}.
\end{equation}
We can rewrite (\ref{Toda}) in terms of $\, \sigma$ given by  (\ref{sigmam}):
\begin{equation}
\label{sigmamother}
\hspace{-0.4in}
  \frac{d}{dt}\ln C_N \, \,= \,\,\,{{\sigma \, +\frac{t}{4} } \over {t \cdot \, (t-1)  }}. 
\end{equation}
Relation (\ref{Toda}) becomes $\, {\cal L} \, = \, {\cal R} \, \, $ where: 
\begin{eqnarray}
\label{Todabecom}
  \hspace{-0.68in}&& \,  \, \quad  \quad  \quad  \, \, \,\, \,
{\cal L} \, = \, \, \,
 t \cdot \, {{d} \over {dt}}  \Bigl( {{\sigma \, +\frac{t}{4} } \over {t \cdot \, (t-1)  }}  \Bigr)
       \,   \,   \, + {{\sigma \, +\frac{t}{4} } \over {t \cdot \, (t-1)  }}
\, \, \, \,  +{{N^2} \over {1 \, -t}^2},
\nonumber \\
  \hspace{-0.68in}&& \, \, \, \, \quad  \quad \quad   \,\, \,
{\cal R}      \,  \, = \, \, \,  \, \,
 {{N^2 \, -1/4} \over {1 \, -t}^2} \cdot \, {{ C_{N-1} \cdot \,  C_{N+1} } \over {C_N^2 }}.
\end{eqnarray}
Let us introduce a new sigma corresponding to the product  $\, C_{N-1} \cdot \,  C_{N+1}$:
\begin{equation}
\label{Sigmam}
\hspace{-0.4in}
\Sigma\, \,= \,\,\,
t \cdot \, (t-1) \cdot \, \frac{d}{dt}\ln \Bigl(C_{N-1} \cdot \,  C_{N+1}\Bigr).
\end{equation}
 Taking a well-suited log-derivatives the previous relation $\, {\cal L} \, = \, \, {\cal R} \, \, $ yields: 
\begin{equation}
\label{log}
\hspace{-0.4in}
t \cdot \, (t-1) \cdot \, \frac{d}{dt}\ln  {\cal L} \, \, \, = \, \,\,  \, t \cdot \, (t-1) \cdot \, \frac{d}{dt}\ln  {\cal R},  
\end{equation}
where the RHS of (\ref{log}) can be written using (\ref{sigmam}) and  (\ref{Sigmam})
\begin{equation}
\label{canbe}
\hspace{-0.4in}
\Sigma \, \, \, \, -2 \, \, \sigma  \,  \,  \, \,  -{{5 \, t} \over {2}}.
\end{equation}
Relation (\ref{log}) becomes:
\begin{eqnarray}
  \label{becomes}
\hspace{-0.98in}&& \,  \, \,  \, \, 
8 \cdot \, t \cdot \, (t-1)^2 \cdot \,  \sigma'' \, \,  \,   +4 \cdot \, (t-1) \cdot \, (t \, +4 \, \sigma) \cdot \,  \sigma'
\, \,  \, -16 \cdot \, \sigma^2 \,  \,  +4 \cdot \, (4\, N^2-1 \,  -t)  \cdot \, \sigma \, 
\nonumber \\
 \hspace{-0.98in}&& \,  \, \, \,    \quad  \quad 
 \,   +(4\, N^2-1) \cdot \, t \, \, 
 -2 \cdot \, \Bigl(4\,N^2-1 \,+4 \cdot \,(t-1) \cdot \,  \sigma' \,  -4 \, \sigma \Bigr) \cdot  \, \Sigma
 \,  \,  \,  \, = \, \,  \, \, 0.
\end{eqnarray}
We can now use the non-linear ODE (\ref{jmequation}) to perform some differential algebra eliminations to eliminate
$\, \sigma$ and its derivatives in order to get a non-linear ODE on $\, \Sigma$. One first eliminates $\,  \sigma''$ between
(\ref{jmequation}) and (\ref{becomes}), getting a (non-linear) relation  between $\, \sigma$, $\, \sigma' $ and $\, \Sigma$. 
Performing a derivation of this relation one gets a relation  between $\, \sigma$, $\, \sigma' $, $\, \sigma'' $, $\, \Sigma$
and $\, \Sigma'$. Again one eliminates $\, \sigma'' $ between this last relation and (\ref{becomes}), getting a relation between
$\, \sigma$, $\, \sigma' $,  $\, \Sigma$ and $\, \Sigma'$. The elimination of  $\, \sigma' $ using a previous relation gives
a relation between $\, \sigma$,  $\, \Sigma$ and $\, \Sigma'$. A new derivation gives a relation between 
$\, \sigma$,  $\, \Sigma$, $\, \Sigma'$ and $\, \Sigma''$. Finally eliminating $\, \sigma$, one gets a non-linear ODE
between $\, \Sigma$, $\, \Sigma'$ and $\, \Sigma''$.  In other words one can obtain a second order
non-linear ODE on $\, \Sigma$, from the Toda-like relation (\ref{Toda}) and the sigma-form of Painlev\'e VI
non-linear ODE (\ref{jmequation}). This non-linear ODE is too large to be given here\footnote[1]{The non-linear ODE emerges from a resultant
  that factors in different spurious terms,  a polynomial in $\, \Sigma$, $\, \Sigma'$ and $\, \Sigma''$ of degree {\em six} in $\, \Sigma''$
and another polynomial in $\, \Sigma$, $\, \Sigma'$ and $\, \Sigma''$ of degree {\em twelve} in $\, \Sigma''$.}. 
However, it is worth noticing that, again, this non-linear ODE has {\em one-parameter} lambda-extension solution.
One may conjecture that this new non-linear ODE has again the (fixed critical point) Painlev\'e property.
This (very large) second order non-linear ODE is {\em not} quadratic in the second derivative
$\, \Sigma''$, in contrast with Okamoto sigma form of Painlev\'e VI equation.
It is of a much higher degree\footnote[2]{Along this second order but higher degree line let us recall~\cite{Sakka}.}. The question
of the reduction of this quite large non-linear ODE to some Okamoto sigma-form of Painlev\'e VI, or more generally to second order
non-linear ODE of the Painlev\'e type~\cite{Cosgrove},
is a (challenging) open question. The transformations required to achieve such reduction to the sigma-form of Painlev\'e VI
will correspond to drastic generalizations\footnote[3]{In the simple case of the reduction of a
  second-order non-Okamoto non linear ODE to an Okamoto sigma form of Painlev\'e VI equation,
  equations (26), (28) in section 2 of~\cite{bmcm}, give some hint of the complexity of such transformations.} 
of the concept of ``folding transformations''~\cite{TOS,CubicQuartic,Kitaev}.

\vskip .1cm 

\subsection{Another non-linear ODE.}
\label{subMangazeev}

If one tries to obtain, more directly, a non-linear ODE on the product of the two diagonal correlation functions
$\, C(N,N) \cdot \, C(N+2,N+2)$, one can also consider the
sigma-form of Painlev\'e VI equation (\ref{jmequation}) together with the definition of sigma (\ref{sigmam})
and the same  equation and definition (\ref{jmequation}) and  (\ref{sigmam}), but for $\, N +2$,
and obtain by differential algebra eliminations
a non-linear ODE on the sum 
\begin{eqnarray}
\label{SigmamBIS}
\hspace{-0.98in}&& \, \quad  \, \,\Sigma\, \,= \,\,\,
t \cdot \, (t-1) \cdot \, \frac{d}{dt}\ln \Bigl(C_{N} \cdot \,  C_{N+2}\Bigr)
\nonumber \\
\hspace{-0.98in}&& \,  \, \, \quad \quad 
\,  \, = \, \,  \, \,
t \cdot \, (t-1) \cdot \, \frac{d}{dt}\ln  C(N,N)
\,    \,    \,  \,  + \, t \cdot \, (t-1) \cdot \, \frac{d}{dt}\ln C(N+2,N+2).
\end{eqnarray}
which is essentially the sum of the two previous sigmas  (equation (\ref{sigmam}) for $\, N$ and for $\, N +2$).
Let us recall (see page 344 in~\cite{BoshernitzanRubel} and~\cite{Moore})
the results on {\em sums} (but also products, compositions, derivatives, integrals, inverses, etc ...)
of differentially algebraic functions, showing that these sums are also  differentially algebraic functions,
and that one also has (see Theorem 2.2 page 345 in~\cite{BoshernitzanRubel}) that the order of the
non-linear ODE for such  sums is less or equal to the sum of the order of the two non-linear ODEs. In our case
(\ref{SigmamBIS}) one expects the order of the non-linear ODE on $\,  \Sigma$ to be less or equal
to $\, 4 \, = \,  2 \, +2\, $ with a prejudice
for the generic upper bound being four.

\vskip .1cm
{\bf Comment:}
We thus have, at first sight, two non-linear ODEs on (\ref{SigmamBIS}):
a very large but second order non-linear ODE obtained by differential algebra eliminations between  (\ref{jmequation})
and (\ref{becomes}), and another one, probably also very large but fourth order non-linear ODE. Both equations
probably have the fixed critical point Painlev\'e property. As far as lambda-extensions are concerned,
we expect the first one to have one-parameter family of power-series analytic at $\, t= \, 0$, when we expect
two-parameters families of power-series analytic at $\, t= \, 0$ (the two lambda parameters for
$\, \sigma(N)$ and $\, \sigma(N +2)$ are, now, independent). Understanding these different non-linear ODEs
occurring on products of two-point correlation functions and their
corresponding lambda extensions remains a challenging work-in-progress task.  

\vskip .1cm
{\bf Remark: Quantum XY chain correlations.}
Along this line, it is worth recalling 
that the emergence of the product  $\, C_{N-1} \cdot \,  C_{N+1}$, or $\, C(N,N) \cdot \, C(N+2,N+2)$, is reminiscent of the
product $\, C(N,N) \cdot \, C(N+1,N+1)$  which is actually  the {\em xx correlation functions of the quantum XY chain in the absence
of a magnetic field}. Actually, for the xx correlations  of the quantum XY chain,
one has (see  (2.45a) and (2.45b) in  Lieb, Schultz and Mattis paper~\cite{SchultzMattis})
the following relations only valid in the absence
of a magnetic field $\, H=\, 0 \, \, $  {\em i.e. precisely} $\, \nu \, = \, -k$:
\begin{eqnarray}
  \label{factodiag1}
  &&\hspace{-.38in} \,\,\,
 < \sigma_0^{x} \, \sigma_{2\, N}^{x}  >  \, \, \, = \, \, \,  \,  C(N, \, N)^2, \quad
  \\
 \label{factodiag2}
   &&\hspace{-.38in} \,\,\, 
 < \sigma_0^{x} \, \sigma_{2\, N\, -1}^{x}  >  \, \,  \, = \, \,  \, \,   C(N, \, N) \cdot \, C(N-1, \, N-1). 
\end{eqnarray}
Again, from the previous results, we have a strong incentive to find the non-linear ODEs
for the  quantum XY chain correlations\footnote[5]{Note that the non-linear ODE for (\ref{factodiag1})
is obviously an Okamoto sigma-form of Painlev\'e VI equation similar to (\ref{sigmam}).} (\ref{factodiag2}).

\vskip .1cm 

More generally we have a strong incentive to find
non-linear ODEs of the Painlev\'e type for various families of two-point correlation functions like the off-diagonal correlations
$\, C(N, \, N+1)$ for which N.Witte showed~\cite{Witte1} the existence of a Garnier system for such  correlations, and, beyond, 
$\, C(N, \, N+2)$, $\, C(N, \, N+3)$,  ... correlations\footnote[1]{The row correlation function $\, C(0, N)$  is a tau-function of a
  Garnier system with five finite singularities, one fixed at the origin: see Corr.1, pg.7 and Eq.(36), pg. 6 of~\cite{Witte2},
  when $\, C(N,N+1)$ is more a component of a related isomonodromic system
  (at least in the description in~\cite{Witte1}). 
 Preliminary studies for the row correlation functions $\, C(0, N)$
  seem to indicate that the corresponding non-linear ODEs are drastically more complicated even if N.Witte showed
  the existence of Garnier systems for these row correlation functions~\cite{Witte2}.}.

\vskip .6cm 

\section{Conclusion.}
\label{Conclusion}

As underlined in the introduction the two-point correlation functions $\, C(M, \, N)$ of the 2D Ising model,
at $\, \nu = \, -k$, can be seen as
D-finite functions solutions of {\em linear} differential equations, but also, in the same time, as 
solutions of {\em non-linear} differential equations of the Painlev\'e type. Around $\, t=0$ the
other solutions of the linear differential equations are formal series with logarithms (see~\cite{Holonomy,Fuchs}). In contrast
 other solutions of the  non-linear differential equations of the Painlev\'e type
are one-parameter families of power series analytic at $\, t=0$. Such  solutions
are called lambda-extensions~\cite{bm}. This paper has tried to provide an illustration of
a set of the remarkable properties and structures of such lambda-extensions
(resp. $\, \alpha$-extensions). The study of non-linear ODEs in the most general framework
may look hopeless for mathematicians.
However, the square Ising model provides a perfect example of the importance of a
{\em selected set of non-linear ODEs}, namely  non-linear ODEs
{\em of the Painlev\'e type}~\cite{Clarkson}, 
and we tried to show that the analysis of some of their solutions, the lambda-extensions,
is clearly a powerful way to describe these selected
non-linear ODEs in a work-in-progress definition of what could be called the ``symmetries'' of these
non-linear ODEs of the Painlev\'e type.

Although Painlev\'e equations were introduced from purely mathematical considerations
their occurrence in so many domains of physics and theoretical physics is compelling. Let us quote
pele mele: particle physics, solid state physics, field theory, lattice statistical mechanics,
statistical physics~\cite{PainlTracy}, integrable PDE's
and their similarity solutions, enumerative combinatorics, 
Random Matrix Theory~\cite{fw,tw}, even Quantum Gravity~\cite{Fokas},
the Ising model being, of course, the perfect play ground for these remarkable non-linear ODEs.
Unfortunately the compelling evidence of the relevance of these selected non-linear
ODEs in physics, is not able to balance the mainstream opinion among pure mathematicians that nothing
interesting can be done on non-linear problems and that even the word ``non-linear'' is
meaningless\footnote[1]{``Using a term like nonlinear science is like referring to the bulk of zoology
  as the study of non-elephant animals.'' Stanislaw Ulam
  (but the citation could be first a citation of Emile Borel ...).}. 
We tried in this paper to show that interesting non-trivial results can be obtained on
 selected non-linear ODEs. 

The exact results sketched in this paper are a strong incentive to get more non-linear ODEs, for instance on the correlation
functions of XY quantum chain in the absence of magnetic field (which corresponds to the product of two Ising two-point
 Ising correlation functions  $\, C(N,N) \cdot \,  C(N+1,N+1)$, but also on many more  two point off-diagonal 
correlation functions  of the 2D Ising like $\,  C(N,N+1)$,  or  $\,  C(N,N+2) $, or  $\,  C(N,N+3)$.

\vskip .3cm 

{\bf Acknowledgments:} One of us (JMM) would like to thank  R. Conte and  I. Dornic 
for many discussions on Painlev\'e equations. We do thank B.M. McCoy for decades of stimulating exchanges on
these problems of Painlev\'e equations on the Ising model, that are such a strong incentive to
find new results on these fascinating questions. One of us (JMM) thanks N. Witte for many Painlev\'e and Garnier systems discussions.

\vskip .6cm 

\vskip .2cm 


\end{document}